\documentclass[nohyper]{JHEP3}

\usepackage{amsmath,amssymb}
\usepackage{graphicx}
\usepackage{cite}

\newcommand{\be}{\begin{equation}}
\newcommand{\ee}{\end{equation}}
\newcommand{\bea}{\begin{eqnarray}}
\newcommand{\eea}{\end{eqnarray}}

\renewcommand{\thefigure}{\Roman{figure}}
\newcommand{\epi}{\epsilon_{i}} 
\newcommand{\efi}{\epsilon_{f_i}}
\newcommand{\esi}{\epsilon_{s_i}} 
\newcommand{\Nif}{{\displaystyle \sum_k}| \hat{A}
\left( \widetilde{N}_i \to L_k h \right )|^2}
\newcommand{\Nis}{{\displaystyle \sum_k}|\hat{A}( \widetilde{N}_i
\to \widetilde{L}_k H )|^2} 
\newcommand{\Nifb}{{\displaystyle \sum_k}
| \hat{A} \left(\widetilde{N}_i \to \bar{L}_k
\bar{h} \right)|^2} 
\newcommand{\Nisb}{{\displaystyle \sum_k}|\hat{A}(
\widetilde{N}_i \to \widetilde{L}_k^{\dag}H^{\dag})|^2}
\newcommand{\Nifbp}{{\displaystyle \sum_{k'}}
| \hat{A} (\widetilde{N}_i \to \bar{L}_{k'}
\bar{h} )|^2} 
\newcommand{\fNi}{{\displaystyle \sum_k}
| \hat{A} (L_k h \to\widetilde{N}_i)|^2} 
\newcommand{\fbNi}{{\displaystyle \sum_k}| \hat{A}
\left(\bar{L}_k \bar{h} \to \widetilde{N}_i \right)|^2}
\newcommand{\sNi}{{\displaystyle \sum_k}
|\hat{A}( \widetilde{L}_k H \to \widetilde{N}_i )|^2}
\newcommand{\sbNi}{{\displaystyle \sum_k}
| \hat{A}( \widetilde{L}_k^{\dag} H^{\dag} \to
\widetilde{N}_i)|^2} 
\newcommand{\Af}{{\displaystyle \sum_k}| A^{f_k}_i|^2} 
\newcommand{\Afp}{{\displaystyle \sum_{k'}}| A^{f_{k'}}_i|^2} 
\newcommand{\As}{{\displaystyle \sum_k}|A^{s_k}_i|^2}
\newcommand{\epm}{| \epsilon_{-} |} \newcommand{\ep}{|\epsilon_{+}|}
\newcommand{\phim}{\phi_{-}} \newcommand{\phip}{\phi_{+}}
 \newcommand{\fh}{f_h^{eq}}
 \newcommand{\fH}{f_H^{eq}}
\newcommand{\fLeq}{f_L^{eq}} 
\newcommand{\fLteq}{f_{\widetilde{L}}^{eq}}

\newcommand{\fNieq}{f_{\widetilde{N}_i}^{eq}}
\newcommand{\ffNi}{f_{\widetilde{N}_i}}
\newcommand{\YNieq}{Y_{\widetilde{N}_i}^{eq}}

\newcommand{\YLf}{Y_{\cal{L}}}
\newcommand{\YLs}{Y_{\widetilde{\cal{L}}}} \newcommand{\YL}{Y_L}
\newcommand{\YLb}{Y_{\bar{L}}} \newcommand{\YLt}{Y_{\widetilde{L}}}
\newcommand{\YLtb}{Y_{\widetilde{L}^{\dag}}}

\title{Soft leptogenesis in the inverse seesaw model}
\author{J. Garayoa\\
Depto.\ de F\'{\i}sica Te\'orica,
and IFIC, Universidad de
Valencia-CSIC \\ 
Edificio de Institutos de Paterna, Apt. 22085, 46071 Valencia,
Spain\\
Email: \email{garayoa@ific.uv.es} }
\author{M.~C.~Gonzalez-Garcia\\
  Instituci\'o Catalana de Recerca i Estudis Avan\c{c}ats (ICREA), \\
  Departament d'Estructura i Constituents de la Mat\`eria,
  Universitat de Barcelona,\\
  Diagonal 647, E-08028 Barcelona, Spain\\
{\rm and:}  C.N. Yang Institute for Theoretical Physics\\
  State University of New York at Stony Brook\\
  Stony Brook, NY 11794-3840, USA,\\
  E-mail: \email{concha@insti.physics.sunysb.edu}}
\author{N. Rius\\
Depto.\ de F\'{\i}sica Te\'orica and
IFIC, Universidad de
Valencia-CSIC \\ 
Edificio de Institutos de Paterna, Apt. 22085, 46071 Valencia,
Spain\\
Email: \email{nuria@ific.uv.es} }

\abstract{
We consider leptogenesis induced by soft supersymmetry breaking terms
(``soft leptogenesis''), in the context of the inverse seesaw
mechanism. In this model there are lepton number ($L$) conserving 
and $L$-violating soft supersymmetry-breaking $B$-terms involving 
the singlet sneutrinos
which, together with the -- generically small-- 
$L$-violating parameter responsible of the neutrino mass, give a small
mass splitting between the four singlet sneutrino states of a single
generation. In combination with the trilinear soft supersymmetry
breaking terms they also provide new CP violating phases needed to
generate a lepton asymmetry in the singlet sneutrino decays.  We
obtain that in this scenario the lepton asymmetry is proportional to
the $L$-conserving soft supersymmetry-breaking $B$-term, and it is not
suppressed by the $L$-violating parameters.
Consequently we find that, as in the standard see-saw case, this
mechanism can lead to sucessful leptogenesis only for relatively small
value of the relevant soft bilinear coupling.  The right-handed
neutrino masses can be sufficiently low to elude the gravitino 
problem. 
Also the corresponding Yukawa couplings involving the lightest
of the right-handed neutrinos are constrained to be 
$\sum |Y_{1k}|^2\lesssim 10^{-7}$  which generically
implies that the neutrino mass spectrum has to be strongly
hierarchical.}
\preprint{%
  UB-ECM-PF-06-37\\
  YITP-SB-06-51\\
  IFIC/06-41 \\
  FTUV-06-1122}

\begin{document}

\section{Introduction}

The discovery of neutrino oscillations makes leptogenesis a very
attractive solution to the baryon asymmetry problem \cite{fy}.  In
the standard framework it is 
usually assumed that the tiny neutrino masses are generated via the
(type I) seesaw mechanism \cite{ss} and thus the new singlet neutral
leptons with heavy (lepton number violating) Majorana masses can
produce dynamically a lepton asymmetry through out of equilibrium
decay. Eventually, this lepton asymmetry is partially converted into a
baryon asymmetry due to fast $B-L$ violating sphaleron processes.

For a hierarchical spectrum of right-handed neutrinos,  
successful leptogenesis requires generically quite heavy singlet neutrino 
masses~\cite{di}, of order $M>2.4 (0.4)\times 10^9$~GeV for vanishing
(thermal) initial neutrino densities~\cite{di,Mbound}, 
although flavour effects \cite{flavour} and/or extended scenarios
\cite{ma} may affect this limit
\footnote{This bound applies when the lepton asymmetry is generated 
in the decay of the lightest right-handed neutrino. The possibility 
to evade the bound producing the asymmetry from the second lightest
right-handed neutrino has been considered in \cite{db1}, and  
flavour effects have been analysed for this case in \cite{oscar}.}.
The stability of
the hierarchy between this new scale and the electroweak one is
natural in low-energy supersymmetry, but in the supersymmetric seesaw
scenario there is some conflict between the gravitino bound on the
reheat temperature and the thermal production of right-handed
neutrinos \cite{gravi}.  This is so because in a high temperature
plasma, gravitinos are copiously produced, and their late decay could
modify the light nuclei abundances, contrary to observation. This sets
an upper bound on the reheat temperature after inflation, $T_{RH} <
10^{8-10}$ GeV, which may be too low for the right-handed neutrinos to
be thermally produced.

Once supersymmetry has been introduced, leptogenesis is induced also
in singlet sneutrino decays.  If supersymmetry is not broken, the
order of magnitude of the asymmetry and the basic mechanism are the
same as in the non-supersymmetric case. However, as shown in 
Refs.\cite{soft1,soft2}, supersymmetry-breaking terms can play an 
important role in the lepton asymmetry generated in sneutrino decays
because they induce effects which are essentially different 
from the neutrino ones.  In brief, soft
supersymmetry-breaking terms involving the singlet sneutrinos remove
the mass degeneracy between the two real sneutrino states of a single
neutrino generation, and provide new sources of lepton number and CP
violation. As a consequence, the mixing between the two sneutrino
states generates a CP asymmetry in the decay, which can be sizable for
a certain range of parameters. In particular the asymmetry is large
for a right-handed neutrino mass scale relatively low, 
in the range $10^{5}-10^{8}$ GeV, well below the reheat temperature limits, 
what solves the cosmological gravitino problem.  Moreover, 
contrary to the traditional leptogenesis
scenario, where at least two generations of right-handed neutrinos are
required to generate a CP asymmetry in neutrino/sneutrino decays, in
this new mechanism for leptogenesis the CP asymmetry in sneutrino
decays is present even if a single generation is considered.  This
scenario has been termed ``soft leptogenesis'', since the soft terms
and not flavour physics provide the necessary mass splitting and
CP-violating phase.  It has also been studied in the minimal
supersymmetric triplet seesaw model \cite{anna}.

In this paper we want to explore soft leptogenesis in the framework of
an alternative mechanism to generate small neutrino masses, namely the
inverse seesaw scheme \cite{iss}. 
This scheme is characterized by a
{\em small} lepton number violating Majorana mass term $\mu$, 
while
the effective light neutrino mass is $m_\nu \propto \mu$. Small values
of $\mu$ are technically natural, given that when $\mu \rightarrow 0$
a larger symmetry is realized \cite{tHooft}: lepton number is
conserved and neutrinos become massless. In the inverse seesaw scheme
lepton flavour and CP violation can arise even in the limit where
lepton number is strictly conserved and the light neutrinos are
massless \cite{pheno1}, due to the mixing of the SU(2) doublet
neutrinos with new SU(2) $\times$ U(1) singlet leptons.

As opposite to the standard seesaw case, these singlet leptons do not
need to be very heavy~\cite{pheno2}, and, as a result, lepton flavour
and CP violating processes are highly enhanced~\cite{pheno1}.  In
Ref.~\cite{hr} it was studied the possibility that the baryon
asymmetry is generated in this type of models during the electroweak
phase transition, in the limit $\mu = 0$. A suppression was found due
to the experimental constraints on the mixing angles of the neutrinos
\cite{pheno3}.  Therefore we consider here the supersymmetric version
of the model and the soft leptogenesis mechanism, since $(i)$ in this
case we expect that a CP asymmetry will be generated in sneutrino
decays even with a single-generation and no suppression due to the
mixing angles is expected, and $(ii)$ this scheme provides a more
natural framework for the relatively low right-handed neutrino mass
scale.

The outline of the paper is as follows.  Section~\ref{issm} presents
the main features of the inverse seesaw model in the presence of
supersymmetry breaking terms.  
In Sec. \ref{qft} we evaluate the
lepton asymmetry generated in the decay of the singlet sneutrinos
using a field-theoretical approach assuming a hierarchy between the
SUSY and $L$ breaking scales and the mass scale of the singlet
sneutrinos.  The relevant Boltzmann equations describing the decay,
inverse decay and scattering processes involving the singlet sneutrino
states are derived in Sec.~\ref{bes}.  Finally in Sec. \ref{res} we
present our quantitative results. In appendix A we recompute the
asymmetry using a quantum mechanics approach, based on an effective
(non hermitic) Hamiltonian \cite{soft1,soft2}.

\section{Inverse Seesaw Mechanism}
\label{issm}

In this type of models \cite{iss}, 
the lepton sector of the 
Standard Model is extended with 
two electroweak singlet two-component leptons per generation, i.e., 
\be
L^i = 
\left( \begin{array}{c}
\nu_L^i \\ 
e_L^i
\end{array} \right),e_R^i, \nu_R^i, s_L^i 
\ee
We assign lepton number $L=1$ to the singlets $s_L^i$ and 
$\nu_R^i$.
In the original formulation of the model, the singlets $s_L^i$ were 
superstring inspired E(6) singlets, in contrast to the right-handed
neutrinos $\nu_R^i$, which are in the spinorial representation. 
More recently this mechanism has also arisen in the context of 
left-right symmetry \cite{alsv} and SO(10) unified models \cite{barr}.

The (9 $\times$ 9) mass matrix of the neutral lepton sector in the 
$\nu_L, \nu_R^c, s_L$ basis is given by 
\be
\label{M}
\cal{M} = 
\left( \begin{array}{ccc}
0    &    m_D    &    0 \\
m_D^{T}&    0    &    M^{T} \\
0    &    M    &    \mu  \\
\end{array} \right)
\ee
where $m_D, M$ are arbitrary 3 $\times$ 3 complex matrices in flavour space 
and $\mu$ is complex symmetric. 
In models where lepton number is spontaneously broken by a vacuum 
expectation value $\langle \sigma \rangle$, 
$\mu = \lambda \langle \sigma \rangle$ \cite{cv1}.
The matrix $\cal{M}$ can be diagonalized 
by a unitary transformation, leading to nine mass eigenstates $n_a$: 
three of them correspond to the observed light neutrinos,
while the other three pairs of two component leptons combine 
to form three quasi-Dirac leptons. 

In this ``inverse seesaw'' scheme, assuming  $m_D,\mu \ll M$
the effective Majorana mass matrix for the light neutrinos is 
approximately given by 
\be
\label{mnu}
m_\nu = m_D^T {M^T}^{-1}\mu M^{-1}m_D \ ,
\ee
while the three pairs of heavy neutrinos have masses of order $M$, and 
the admixture among singlet and doublet $SU(2)$ states is 
suppressed by $m_D/M$. 
Although $M$ is a large mass scale suppressing the light neutrino masses,
in contrast to the Majorana mass ($\Delta L =2$) of the right-handed neutrinos 
in the standard seesaw mechanism, it is a Dirac mass 
($\Delta L = 0$), and it can be much smaller, since
the suppression in Eq.~(\ref{mnu}) is quadratic and moreover light 
neutrino masses are 
further suppressed by the small parameter $\mu$ which characterizes
the lepton  number violation scale. 

Notice that in the $\mu \rightarrow 0$ limit lepton number conservation is 
restored. Then, 
the three light neutrinos are massless Weyl particles and the six 
heavy neutral leptons combine exactly into three Dirac fermions. 

We are going to consider the supersymmetric version of this model
(for an analysis of lepton flavour violation in this case see \cite{siss}).  
In this case, the above neutral lepton mass matrix (\ref{M}) is described 
by the following superpotential:
\be 
\label{W} 
W = Y_{ij} N_i L_j  H + \frac{1}{2}\mu_{ij} S_i S_j + M_{ij} S_i N_j   \ ,
\ee
where $i,j =1,2,3$ are flavour indices, $H,L_i,N_i,S_i$ are the 
superfields corresponding to the $SU(2)$ 
up-Higgs and lepton doublets, 
and $\nu_R^{ic}$ and $s_L^i$ singlets, respectively, 
and $Y_{ij}$ denote the neutrino Yukawa couplings. Thus, after 
spontaneous electroweak symmetry breaking, the neutrino Dirac masses are
given by
\be
(m_D)_{ij}= Y_{ij} \langle H \rangle
\ee
The relevant soft supersymmetry breaking terms are the bilinear and
trilinear scalar couplings involving the singlet sneutrino fields, 
that provide new sources of lepton number and CP violation.
From now on we consider a simplified one generation of $(N,S)$ model
because a single generation of singlet sneutrinos is sufficient to
generate the CP asymmetry. Indeed in the three-generation case, the relevant
out of equilibrium decays are usually those of the lightest heavy
singlet states while the decay of the heavier (if heavier enough) 
give no effect. Thus with our simplified single generation model
we refer to the lightest of the three heavy singlet sneutrinos which
we number as $1$. Consequently we label $M=M_{11}$ and $\mu=\mu_{11}$. 
Also, for simplicity, we will assume proportionality of the soft trilinear
terms.
\be  
\label{soft} 
- L_{soft} = A Y_{1i} \widetilde{L}_i \widetilde{N} H + 
\widetilde{m}_S^{2}\widetilde{S}\widetilde{S}^{\dag} + 
\widetilde{m}_N^{2}\widetilde{N}\widetilde{N}^{\dag} +  
\widetilde{m}_{SN}^2 \widetilde{S}\widetilde{N}^{\dag} +
B_S \widetilde{S} \widetilde{S} + 
B_{SN} \widetilde{S} \widetilde{N} + h.c. 
\ee
With our lepton number assignments, the soft SUSY breaking terms 
which violate L are $\widetilde{m}_{SN}^2$ and $B_S$.
The sneutrino interaction Lagrangian is then:
\begin{eqnarray} 
\label{L}
\lefteqn{ L = 
(Y_{1i} L_i N H + Y_{1i} L_i \widetilde{N} h + Y_{1i} \widetilde{L}_i 
N h + h.c.)  + {}  }  
\nonumber\\
& & {} + 
( Y_{1i} M^{*} \widetilde{L}_i \widetilde{S}^{\dag} H +  A Y_{1i} 
\widetilde{L}_i 
\widetilde{N} H + h.c.) + {} \nonumber\\
& & {} +(\mu SS + M S N + h.c.) +  {} \nonumber\\
& & {} +((\left|\mu\right|^{2}+\widetilde{m}_S^{2}+\left| M\right|^{2}) 
\widetilde{S}^{\dag} \widetilde{S} 
+( \widetilde{m}_N^{2}+\left| M\right|^{2})\widetilde{N}\widetilde{N}^{\dag} 
+(\mu M^{*}+\widetilde{m}_{SN}^2) \widetilde{S} \widetilde{N}^{\dag} + h.c.)
+ {} \nonumber\\
& & {} + (B_S \widetilde{S} \widetilde{S} + 
B_{SN} \widetilde{S} \widetilde{N} + h.c.)  \end{eqnarray}
This Lagrangian  has three independent physical CP violating phases: 
$\phi_B$ which can be assigned  to $B_{SN}$, $\phi_A$  which 
is common to the three   terms with $A Y_{1i}$, and  $\phi_{MY}$ which 
is common to theT three terms with  $M Y^*_{1i}$, and are given by:
\be 
\label{CPphases}
\begin{array}{c}
\phi_B = \arg(B_{SN} B_S^{*} \widetilde{M}_{SN}^2)  \\
\phi_A = \arg(A B_S^{*} M^2 \mu^* (\widetilde{M}_{SN}^2)^2)  \\
\phi_{MY} = \arg(\widetilde{M}_{SN}^{2*} M^* \mu) \ ,       \\
\end{array}  
\ee
where we have defined 
$\widetilde{M}_{SN}^2 \equiv \mu M^* + \widetilde{m}_{SN}^2$.
These phases provide the CP violation necessary to generate dynamically 
a lepton asymmetry, even with a single generation of sneutrinos.
They can also contribute to lepton electric dipole moments~\cite{farzan}.

From the Lagrangian in Eq.~(\ref{L}) we obtain the sneutrino mass matrix 
in the interaction basis  ($\widetilde{F}_i\equiv 
\widetilde{N},\widetilde{N}^{\dag}, \widetilde{S},\widetilde{S}^{\dag}$):
\be 
\label{msn}
\left( \begin{array}{cccc} 
\widetilde{m}_N^{2}+\left| M\right|^{2} & 0 &
\widetilde{M}_{SN}^{2*} & B_{SN} \\ 
0 & \widetilde{m}_N^{2}+\left| M\right|^{2} & B_{SN}^* & 
\widetilde{M}_{SN}^2 \\ 
\widetilde{M}_{SN}^2 & B_{SN} & 
\left|\mu\right|^{2}+\widetilde{m}_S^{2}+\left| M\right|^{2}
& 2B_S \\ B_{SN}^* & \widetilde{M}_{SN}^{2*} &
2B_S^{*} & \left|\mu\right|^{2}+\widetilde{m}_S^{2}+\left|
M\right|^{2} \\
\end{array} \right) 
\ee 
Notice that in the most general case it is not possible to remove all CP 
phases  from the sneutrino mass matrix. With our choice of 
basis, $B_{SN}$ is the only complex parameter, 
$B_{SN} = |B_{SN}| e^{i\phi_B}$.

Although one can easily obtain the analytic expressions for the corresponding 
mass  eigenvalues and eigenvectors, for the general case they are
lengthy and we do not give them here. 
Under the assumption that all the entries are real, i.e., $\phi_B =0$,   
we obtain the following mass eigenvalues:
\bea 
 M_1^2 &=& M'^2 + B_S - \frac 1 2 \sqrt{4(B_{SN} + \widetilde{M}_{SN}^{2})^2 +
(2 B_S - \widetilde{m}_N^{2} + \widetilde{m}_S^{2} + |\mu|^2)^2}
     \nonumber\\
 M_2^2&=& M'^2 - B_S + \frac 1 2 \sqrt{4(B_{SN} - \widetilde{M}_{SN}^{2})^2 +
(2 B_S + \widetilde{m}_N^{2} - \widetilde{m}_S^{2} - |\mu|^2)^2}
\nonumber\\
 M_3^2&=&M'^2 - B_S - \frac 1 2 \sqrt{4(B_{SN} - \widetilde{M}_{SN}^{2})^2 +
(2 B_S + \widetilde{m}_N^{2} - \widetilde{m}_S^{2} - |\mu|^2)^2}
\nonumber \\
M_4^2&=& M'^2 + B_S + \frac 1 2 \sqrt{4(B_{SN} + \widetilde{M}_{SN}^{2})^2 +
(2 B_S - \widetilde{m}_N^{2} + \widetilde{m}_S^{2} + |\mu|^2)^2} \ ,
\label{full}
\eea
where we have defined 
$M'^2 \equiv  |M|^2 + \widetilde{m}_N^{2} + \widetilde{m}_S^{2} + |\mu|^2$.

Furthermore, if  we assume conservative values of the soft breaking terms:
\be 
\begin{array}{c}
A \sim {\cal O}(m_{SUSY}) \\
\widetilde{m}_{N} \sim \widetilde{m}_S \sim \widetilde{m}_{SN} 
\sim {\cal O}(m_{SUSY}) 
\\
B_S \sim {\cal O}(m_{SUSY} \mu)  \\
B_{SN}\sim {\cal O}(m_{SUSY} M)  
\label{estimates}
\end{array}
\ee
with both, $\mu, m_{SUSY} \ll M$, we see that 
$B_S, \widetilde{m}_N^2, \widetilde{m}_S^2, \widetilde{m}_{SN}^2  \ll B_{SN}$
and $\widetilde{M}_{SN}^2 \sim \mu M^*$.  
Neglecting these small soft terms, there is still one physical CP violating 
phase, 
\be
\label{CP1}
\phi= \phi_A - \phi_B = \arg(A B_{SN}^* M) \ .
\ee
In this limit, we choose for simplicity a basis where 
$A=|A| e^{i \phi}$ is the only complex parameter. Then 
we diagonalize to first order in two expansion parameters, 
\be
\label{expanp}
\epsilon = \frac{|\mu|}{2|M|} \ ,
\hspace{1.cm} 
\tilde{\epsilon}=\frac{|B_{SN}|}{2 |M|^2} \sim {\cal O}(m_{SUSY}/M) 
\ee
To this order the mass eigenvalues are:
\bea 
 M_1^2 &=& M^2- M\mu-B_{SN} \nonumber\\
 M_2^2&=&M^2-M\mu+B_{SN} \nonumber\\
 M_3^2&=&M^2+ M\mu-B_{SN} \label{masses}
\\
 M_4^2&=&M^2+M\mu+B_{SN} \nonumber
\eea
and the eigenvectors:
\bea
\widetilde{N}_1&=& \frac{1}{2}\left( \widetilde{S}^{\dag}
-\widetilde{N}^{\dag}
\right)+\frac{1}{2}\left(\widetilde{S}-\widetilde{N}\right) \nonumber \\
\widetilde{N}_2&=& \frac{i}{2}\left( \widetilde{S}^{\dag}
-\widetilde{N}^{\dag}
\right)-\frac{i}{2}\left(\widetilde{S}-\widetilde{N}\right)  \nonumber \\
\widetilde{N}_3 &=&\frac{i}{2}\left( \widetilde{S}^{\dag}
+\widetilde{N}^{\dag}
\right)-\frac{i}{2}\left(\widetilde{S}+\widetilde{N}\right) 
\label{masseigenstates}
\\
\widetilde{N}_4&=&\frac{1}{2}\left( \widetilde{S}^{\dag}
+\widetilde{N}^{\dag}
\right)+\frac{1}{2}\left(\widetilde{S}+\widetilde{N}\right) \nonumber
\eea
Note that in this limit the mass degeneracy among the four 
sneutrino states is removed by both the $L$-violating mass $\mu$ and 
$L$-conserving supersymmetry breaking term $B_{SN}$. 
Together with the trilinear $A$ term they also provide a source of CP
violation, and the mixing among the four sneutrino states leads to
a CP asymmetry in their decay.

Another interesting limit is to diagonalize the sneutrino mass matrix 
(\ref{msn}) neglecting only the $B_S$ entry, which may be appropriate if 
$\mu \ll m_{SUSY}$ and the order of magnitude of the soft breaking terms 
is as given by (\ref{estimates}).
In this limit the mass matrix can also be taken real and
the mass eigenvalues can be read from Eq. (\ref{full}), just setting 
$B_S =0$. Now there are two non zero CP violating phases, 
$\phi_{YM}$ and 
$\phi'_A = \arg (A B_{SN}^{*}  M^2 \mu^* \widetilde{M}_{SN}^2)$. 
However the combination that is relevant for the
CP asymmetry in sneutrino decays is the same as in the previous case, 
$\phi= \phi_{YM} + \phi'_A=\arg(A B_{SN}^* M) $.

As we will see in Sec.~\ref{res}, the total CP asymmetry in the singlet 
sneutrino decays
turns out to be sizable for very small values of the soft term
$B_{SN} \ll M m_{SUSY}$. Neglecting the $B_{SN}$ term in the
Lagrangian there are still two CP violating phases, $\phi_{MY}$ and
$\phi_A$, but again the sneutrino mass matrix can be taken real, so
that the mass eigenvalues are as given by Eq.(\ref{full}) with
$B_{SN}=0$. The phase relevant for the CP asymmetry in the singlet
sneutrino decays is now $\phi' = \phi_{YM} + \phi_A = \arg(A B_S^{*} M
\widetilde{M}_{SN}^2)$. 

Finally, neglecting supersymmetry breaking effects,  
the total singlet sneutrino decay width is given by  
\be
\Gamma = \frac {\displaystyle \sum_i |M| |Y_{1i}|^2}{\displaystyle 8 \pi} \ .
\label{eq:gamma}
\ee

\section{The CP Asymmetry}
\label{qft}

In this section we compute the CP asymmetry in the singlet sneutrino decays.
As discussed in Ref.\cite{soft2}, 
when $\Gamma \gg \Delta M_{ij}\equiv M_i -M_j $, the four 
singlet sneutrino states are not well-separated 
particles. In this case, the result for the asymmetry depends on how the
initial state is prepared.
In what follows we will assume that 
the sneutrinos are in a thermal 
bath with a thermalization time $\Gamma^{-1}$ shorter than the 
typical oscillation times, $\Delta M_{ij}^{-1}$, therefore coherence 
is lost and it is appropriate to compute the CP asymmetry  in terms of the 
mass eigenstates Eq.(\ref{masseigenstates}).

The CP asymmetry produced in the decay of the state $\widetilde{N}_i$
is given by (see sec.\ref{bes}):
\be
\label{epi}
\epi = \frac{\displaystyle \sum_f \Gamma(\widetilde{N}_i \rightarrow f)
- \Gamma(\widetilde{N}_i \rightarrow \bar{f})}
{\displaystyle \sum_f \Gamma(\widetilde{N}_i \rightarrow f)
+ \Gamma(\widetilde{N}_i \rightarrow \bar{f})} \ , 
\ee
where $f=\widetilde{L}_k H, L_k h$. 
We also define the fermionic and scalar CP asymmetries in the decay of each 
$\widetilde{N}_i$ as
\begin{eqnarray} 
\esi & = &\frac{\displaystyle \sum_k |\hat{A}_i
(\widetilde{N}_i \to \widetilde{L}_k H )|^2 -
|\hat{A}_i(\widetilde{N}_i  \to  \widetilde{L}_k^{\dag} H^{\dag})|^2}
{\displaystyle \sum_k|\hat{A}_i(\widetilde{N}_i \to \widetilde{L}_k H )|^2 +
|\hat{A}_i(\widetilde{N}_i  \to  \widetilde{L}_k^{\dag} H^{\dag})|^2}
\\ 
\efi & = & \frac {\displaystyle \sum_k
|\hat{A}_i(\widetilde{N}_i \to L_k h  )|^2 -
|\hat{A}_i(\widetilde{N}_i  \to  \bar{L}_k \bar{h})|^2}
{\displaystyle \sum_k |\hat{A}_i(\widetilde{N}_i \to L_k h  )|^2 +
|\hat{A}_i(\widetilde{N}_i  \to  \bar{L_k} \bar{h})|^2}
 \label{asymdefhat} \ .
\end{eqnarray}
Notice that $\esi$ and $\efi$ are defined in terms of decay amplitudes,
without the phase-space factors which, as we will see, are crucial to
obtain a non-vanishing CP asymmetry, much as in the standard seesaw
case \cite{soft1,soft2}. 
The total asymmetry $\epi$ generated in the decay of the singlet sneutrino
$\tilde N_i$ can then be written as \be
\epi = \frac{\esi c_s + \efi c_f}{c_s + c_f}  \ ,
\ee
where $c_s, c_f$ are the phase-space factors of the scalar and 
fermionic channels, respectively.  

Since the scale of lepton number and supersymmetry breaking are
$\mu, m_{SUSY} \ll M$, there is an enhancement of the CP violation 
in mixing (wave-function diagrams), so we only include this leading
effect and neglect direct CP violation in the decay (vertex diagrams). 

We compute the CP asymmetry 
following the effective field theory approach described in \cite{pi},
which takes into account the CP violation 
due to mixing of nearly degenerate states by using
resumed propagators for unstable (mass eigenstate) particles.
The decay amplitude $\hat{A}_i^f$ of the unstable external state 
$\widetilde{N}_i$ defined in Eq.~(\ref{masseigenstates}) into a final
state $f$ is described by a superposition of amplitudes
with stable final states:
\be
\label{resum}
\hat{A}_i(\widetilde{N}_i \rightarrow f) = A_i^f -  \sum_{j\neq i} A_j^f 
\frac{i \Pi_{ij}}{M_i^2 - M_j^2 + i \Pi_{jj}} \ , 
\ee
where $A_i^f$ are the tree level decay amplitudes
and $\Pi_{ij}$ are the absorptive parts of the two-point functions
for $i,j=1,2,3,4$.
The amplitude for the decay into the conjugate final state is 
obtained from (\ref{resum}) by the replacement $A_i^f \rightarrow 
A_i^{f*}$.

The decay amplitudes can be read off from the interaction Lagrangian 
(\ref{L}), after performing the change from the current 
to the mass eigenstate basis:
\bea
\label{L2}
-{\cal L} &=& 
\frac 1 2 \widetilde{N}_1 [- Y_{1k} L_k h + (Y_{1k} M^{*} - A Y_{1k}) 
\widetilde{L}_k H]
\nonumber \\
&+& \frac {i}{ 2} \widetilde{N}_2 [- Y_{1k} L_k h - (Y_{1k} M^{*} + A Y_{1k}) 
\widetilde{L}_kH]
\nonumber \\
&+& \frac {i}{ 2} \widetilde{N}_3 [Y_{1k} L_k h - (Y_{1k} M^{*} - A Y_{1k}) 
\widetilde{L}_kH]
\nonumber \\
&+&
\frac 1 2 \widetilde{N}_4 [Y_{1k} L_k h + 
(Y_{1k} M^{*} + A Y_{1k}) \widetilde{L}_kH] + h.c.
\eea

Neglecting supersymmetry breaking in vertices, up to an overall normalization
we obtain  that the decay amplitudes into scalars,
$A^{s_k}_i=A(\widetilde{N}_i \to \widetilde{L}_k H  )$, verify 
$A_2^{s_k}=A_3^{s_k}=i A_1^{s_k}= i A_4^s$. 
Correspondingly the decay amplitudes into fermions 
$A^{f_k}_i=A(\widetilde{N}_i \to L_k h  )$, verify
$A_2^{f_k}=-A_3^{f_k}=-i A_1^{f_k}=i A_4^{f_k} $.  

Keeping only the lowest order contribution in the soft terms, 
\bea
\label{2pa}
\Pi_{ii} &=& M\, \Gamma \quad i=1,\ldots,4 \\
\Pi_{12} & = & \Pi_{21} = - \Pi_{34} = - \Pi_{43}= 
|A| \, \Gamma\,  \sin\phi
\eea
Altogether we can then write 
the fermionic and scalar CP asymmetries as:
\begin{eqnarray} 
\esi & = &
\sum_{j \neq i}
\frac{\displaystyle 2 (M_i^2 - M_j^2) 
\Pi_{ji}\sum_k {\rm Im}(A_i^{s_k*} A_j^{s_k})}
{\displaystyle 
[(M_i^2 - M_j^2)^2 + \Pi_{jj}^2] \sum_k |A_i^{s_k}|^2}
\\ 
\efi & = &  
\sum_{j \neq i}
\frac {\displaystyle 2 (M_i^2 - M_j^2)\Pi_{ji}
\sum_k {\rm Im}(A_i^{f_k*} A_j^{f_k})}
{\displaystyle[(M_i^2 - M_j^2)^2 + \Pi_{jj}^2] \sum_k|A_i^{f_k}|^2} 
 \label{asymhat}
\end{eqnarray}
Inserting the values of the amplitudes $A_i^f$ and the absorptive parts 
of the two-point functions (\ref{2pa}) we obtain the final expression
for the scalar and fermionic CP asymmetries at $T=0$:
\be
\esi= - \efi = \bar{\epsilon}_i =
- \frac{4\, |B_{SN}\, A|\, \Gamma }{4 |B_{SN}|^2 + |M|^2 \Gamma^2} \sin\phi
\label{qftasym}
\ee
and the total CP asymmetry generated in the decay of the sneutrino 
$\widetilde{N}_i$ is then:
\be
\epi (T) = \bar{\epsilon}_i \ \frac{c_s - c_f}{c_s + c_f} \ .
\label{asymi}
\ee
As long as we neglect the zero temperature lepton and slepton masses
and  small Yukawa couplings,  
the phase-space factors of the final states are flavour independent.  
After  including finite temperature effects they are given by:
\bea
c_f&=&(1-x_L -x_h)\lambda(1,x_L,x_h)
\left[ 1-\fLeq\right] \left[ 1-\fh\right] 
\label{cfeq}\\
c_s&=&\lambda(1,x_H,x_{\widetilde{L}})
\left[ 1+\fH\right] \left[ 1+\fLteq\right]
\label{cbeq}
\eea
where
\bea
f^{eq}_{H,\widetilde{L}}&=&\frac{1}{\exp[E_{H,\widetilde{L}}/T]-1}
\label{eq:fHeq}\\
f^{eq}_{h,L}&=& \frac{1}{\exp[E_{h,L}/T]+1}  
\label{eq:fheq}
\eea
are the Bose-Einstein and Fermi-Dirac equilibrium distributions,
respectively, and 
\bea
&E_{L,h}=\frac{M}{2} (1+x_{L,h}-
x_{h,L}), ~~~
E_{H ,\widetilde{L}}=\frac{M}{2} (1+x_{H ,\widetilde{L}}-
x_{\widetilde{L},H})&\\
&\lambda(1,x,y)=\sqrt{(1+x-y)^2-4x},~~~
x_a\equiv \frac{m_a(T)^2}{M^2}&
\eea
The thermal masses for the relevant supersymmetric degrees of
freedom are \cite{thermal}:
\bea
m_H^2(T)=2 m_{h}^2(T)&=& \left(\frac{3}{8}g_2^2+\frac{1}{8}g_Y^2
+\frac{3}{4}\lambda_t^2\right) \, T^2\; ,\\
m_{\widetilde{L}}^2(T)=2 m_L^2(T)&=& \left(\frac{3}{8}g_2^2+\frac{1}{8}g_Y^2
\right)\, T^2\; .
\eea
Here $g_2$ and $g_Y$ are gauge couplings and $\lambda_t$ is the top Yukawa,
renormalized at the appropriate high-energy scale. 

As we will see in the next section, from Eq.~(\ref{eq:ltotal}), 
if the initial distributions of all four states $\tilde N_i$ are equal, their 
total contribution to the total lepton number can be factorized as:
\be
\epsilon(T) \equiv \sum_i \frac
{\displaystyle \sum_{f} \Gamma(\widetilde{N}_i \rightarrow f)
- \Gamma(\widetilde{N}_i \rightarrow \bar{f})}
{\displaystyle \sum_{f} \Gamma(\widetilde{N}_i \rightarrow f)
+ \Gamma(\widetilde{N}_i \rightarrow \bar{f})} 
 \ . 
\ee

Several comments are in order.  We find that this leptogenesis
scenario presents many features analogous to soft leptogenesis in
seesaw models \cite{soft1,soft2,anna}: $(i)$ The CP asymmetry
(\ref{asymi}) vanishes if $c_s = c_f$, because then there is an exact
cancellation between the asymmetry in the fermionic and bosonic
channels. Finite temperature effects break supersymmetry and make the
fermion and boson phase-spaces different $c_s \neq c_f$, mainly
because of the final state Fermi blocking and Bose stimulation
factors.  $(ii)$ It also displays a resonance behaviour: the maximum
value of the asymmetry is obtained for $2 B_{SN}/M \sim \Gamma$.
$(iii)$ The CP asymmetry is due to the presence of supersymmetry
breaking and irremovable CP violating phases, thus it is proportional
to $|B_{SN}\, A| \sin\phi$.

As seen from Eq.(\ref{qftasym}) we obtain that the CP asymmetry is not 
suppressed by the lepton number violating scale $\mu$. This may seem 
counterintuitive. However if $\mu
= 0$ the four sneutrino states are pair degenerate, and we can choose
a lepton number conserving mass basis, made of the ($L=1$) states
\bea
\widetilde{N}'_1&=& \frac{1}{\sqrt{2}}\left( \widetilde{S}^{\dag}
-\widetilde{N} \right) \nonumber \\
\widetilde{N}'_2&=& \frac{1}{\sqrt{2}}\left( \widetilde{S}^{\dag}
+\widetilde{N}\right)
\eea
and their hermitian conjugates, with $L=-1$, 
$\widetilde{N}'^{\dag}_1,\widetilde{N}'^{\dag}_2$. 
Although there is a CP asymmetry in the decay of these sneutrinos, 
it is not a lepton number asymmetry (since in the limit $\mu=0$ total
lepton number is conserved) but just a redistribution of the
lepton number stored in heavy sneutrinos and light lepton
and slepton $SU(2)$ doublets. 
At very low temperatures, $T \ll M$, when no heavy sneutrinos 
remain in the thermal bath, all lepton number is in the light 
species and obviously if we started in a symmetric Universe
with no lepton number asymmetry it will not be generated.

In other words, the total lepton number generated for the case with no 
lepton number violation is zero but it cannot be recovered by taking the limit
$\mu\rightarrow 0$ of Eq.(\ref{qftasym}) because in the derivation of
Eq.(\ref{qftasym}) it is assumed implicitly that the four singlet
sneutrino states are non-degenerate and consequently it is only valid
if $\mu$ (or some of the other $L$ violating parameters) is non zero.

In appendix A we recompute the asymmetry using a quantum mechanics
approach, based on an effective (non hermitic) Hamiltonian
\cite{soft1,soft2}, and we get the same parametric dependence of the
result, which differs only by numerical factors. Both expressions
agree in the limit $\Gamma \ll |B_{SN}/M|$.

As discussed at the end of the previous section there may be other
interesting ranges of parameters beyond Eq.(\ref{expanp}).  Thus in
order to verify the stability of the results to departures from this
expansion we have redone the computation of the CP asymmetry keeping
all the entries in the sneutrino mass matrix, and just assuming that
it is real. The expressions are too lengthy to be given here but let
us simply mention that we have found that, in the general case, the CP
asymmetries generated in the decay of each of the four singlet sneutrino
states are not equal but they can always be written as:
\begin{equation}
\epi=
- \frac{4\, |B_{SN}\, A|\, \Gamma }{4 |B_{SN}|^2 + |M|^2 \Gamma^2} 
\sin\phi
+f_i(B_S,\mu,\widetilde M^2_{SN}) \,
\label{eq:epsigen}
\end{equation}
where the functions $f_i$ verify that for $|\mu|^2\ll |B_{SN}|$ 
and to any order in $|B_S|$: 
\begin{equation}
\sum_i f_i(B_S,\mu,\widetilde{M}^2_{SN}) \propto |B_{SN}| \, .
\label{eq:fis}
\end{equation} 

In the limiting case $|B_{SN}|\ll |B_S|,m_{SUSY}^2,|\mu^2|$
the dominant term in the CP asymmetry at leading order in 
$|B_S| \sim |M| \Gamma \ll |\widetilde{M}_{SN}^2|$
is 
\begin{equation}
\sum_i\epi= \frac{8 |B_S A| \Gamma}
{(4 |B_{S}|^2 + |M|^2 \Gamma^2)^2} 
\frac {|\mu^2| + \widetilde{m}_S^{2} - \widetilde{m}_N^{2}}
{|\widetilde{M}_{SN}^2|} 
(4 |B_{S}|^2 - |M|^2 \Gamma^2) \sin\phi' 
\, .
\label{eq:epsibsn0}
\end{equation}
It also exhibits  a resonant behaviour, described now by  
$|B_S| \Gamma/(4 |B_{S}|^2 + |M|^2 \Gamma^2)^2$, however
the total CP asymmetry in this limit is further suppressed by a 
factor of order $(|\mu^2|,m_{SUSY}^2)/|\widetilde{M}_{SN}^2|$.

Finally, let's comment that in the previous derivation 
we have neglected thermal corrections to the CP asymmetry from the loops, 
i.e., we have computed the imaginary part of the one-loop graphs using 
Cutkosky cutting rules at $T=0$. These corrections 
are the same for scalar and fermion decay channels,
since only bosonic loops contribute to the wave-function diagrams
in both cases, so they are not expected to introduce significant changes to
our results.

\section{Boltzmann Equations}
\label{bes}

We next write the relevant Boltzman equations describing the decay, inverse 
decay and scattering processes involving the sneutrino states.  

As mentioned above we assume that the sneutrinos are in a thermal bath with a 
thermalization time shorter than the oscillation time. Under this assumption
the initial states can be taken as being the mass eigenstates in 
Eq.~(\ref{masseigenstates}) and we write the corresponding equations for
those states and the scalar and fermion lepton numbers. 
The $CP$ fermionic and scalar asymmetries for each $\widetilde{N}_i$ defined
at $T=0$ are those given in Eq.~(\ref{qftasym}).

Let's notice that the CP asymmetries as defined in Eq.~(\ref{qftasym})
verify $\esi=-\efi\equiv\bar\epsilon_i$. However in order to better  
trace the evolution of the scalar and fermion lepton numbers 
separately we will keep them as two different quantities 
in writing the equations.

Using $CPT$ invariance and the above definitions for the $CP$ 
asymmetries and including all the multiplicative factors we have:
\be
\begin{array}{ccccc}
\Nis & = & \sbNi & \simeq &\displaystyle{\frac{1+\esi}{2}\As}  \; ,\\
\Nisb & = & \sNi & \simeq &\displaystyle{\frac{1-\esi}{2}\As} \; , \\
\Nif & = & \fbNi & \simeq &  \displaystyle{\frac{1+\efi}{2}\Af}  \; , \\
\Nifb & = & \fNi & \simeq &\displaystyle{\frac{1-\efi}{2}\Af}  \; . \\
\end{array} 
\ee
where 
\begin{eqnarray}
\As&=&\displaystyle {\sum_k \frac{|Y_{1k} M|^2}{4}} \; , \nonumber \\
\Af&=&\displaystyle {\sum_k \frac{|Y_{1k} M|^2}{4} 
\frac{M_{i}^2}{M^2}} \; . 
\label{amplidef}
\end{eqnarray}
The Boltzmann equations describe the evolution of the number density 
of particles in the plasma:
\begin{eqnarray}
  \frac{dn_X}{dt} + 3Hn_X & = & \sum_{j,l,m} 
\Lambda^{X j \dots}_{l m \dots} \left[ f_l f_m \dots (1\pm f_X) (1\pm f_j)
    \dots W(l m \dots \to X j \dots)   - \right. \nonumber \\
 & - & \left.   f_X f_j \dots (1\pm
  f_l)(1\pm f_m) \dots W(X j \dots \to l m \dots) \right] \nonumber
\end{eqnarray}
where, 
\bea
\Lambda^{X j \dots}_{l m \dots} & = & 
\int \frac{d^3 p_X}{(2\pi)^3 2E_X} 
\int \frac{d^3 p_j}{(2\pi)^3 2E_j} \dots 
\int \frac{d^3 p_l}{(2\pi)^3 2E_l}
\int \frac{d^3 p_m}{(2\pi)^3 2E_m} \dots  \; ,\nonumber
\eea
and $W(l m \dots \to X j \dots)$ is the squared transition amplitude summed
over initial and final spins. 
In what follows we will use the notation of Ref.\cite{kolb} and 
we will assume that the Higgs and higgsino fields are in 
thermal equilibrium with distributions given in 
Eqs.~(\ref{eq:fHeq}) and ~(\ref{eq:fheq}) respectively, while 
the leptons and sleptons are in kinetic
equilibrium and we introduce a chemical potential for the 
leptons, $\mu_f$, and sleptons, $\mu_s$:
\begin{eqnarray}
f_{L}&=&\frac{1}{\exp[(E_L+\mu_f)/T]+1}  \; , \nonumber \\
f_{\bar{L}}&=&\frac{1}{\exp[(E_L-\mu_f)/T]+1} \;,\label{fdist}  \\
f_{\widetilde{L}}&=&
\frac{1}{\exp[(E_{\widetilde L}+\mu_s)/T]-1}  \; , \nonumber 
\\
f_{\widetilde{L}^\dagger}&=&
\frac{1}{\exp[(E_{\widetilde{L}}-\mu_s)/T]-1} \;  . \nonumber  
\end{eqnarray}
Furthermore in order to eliminate the dependence in the expansion of the
Universe we write the equations in terms of the abundances 
$Y_X$, where $Y_X=n_X/s$.

We are interested in the evolution of sneutrinos
$Y_{\widetilde{N}_i}$, and the fermionic $\YLf$ and scalar $\YLs$
lepton number, defined as $\YLf=(\YL-\YLb)/2$,
$\YLs=(\YLt-\YLtb)/2$. The number density of sneutrinos is regulated
through its decays and inverse decays, defined in the $D-terms$, while
to compute the evolution of the fermionic and scalar lepton number we
also need to consider the scatterings where leptons and sleptons are
involved. The scattering terms are defined in the $S-terms$.
\begin{eqnarray}
\frac{d Y_{\widetilde{N}_i}}{d t} & = &
-D_i-\bar{D}_i-\widetilde{D}_i-\widetilde{D}_i^{\dag} 
\label{eq:bolN}\\
\frac{d\YLf}{dt} & = & \sum_i\left( D_i-\bar{D}_i \right)-2 S -
S_{L\widetilde{L}^{\dag}} + \bar{S}_{L\widetilde{L}^{\dag}} -
S_{L\widetilde{L}} + \bar{S}_{L\widetilde{L}} ç
\label{eq:bolLf}\\
\frac{d\YLs}{dt} & = & \sum_i\left( \widetilde{D}_i-\widetilde{D}_i^{\dag} 
\right)-2
\widetilde{S} - S_{L\widetilde{L}^{\dag}} +
\bar{S}_{L\widetilde{L}^{\dag}} + S_{L\widetilde{L}} -
\bar{S}_{L\widetilde{L}}
\label{eq:bolLs}
\end{eqnarray}
where 
\begin{eqnarray}
s D_i & = & \Lambda^{12}_{\widetilde{N}_i}
             \left[f_{\widetilde{N}_i} (1-f_L)(1-\fh)\Nif -
             \right. \nonumber \\ 
               & & \left. - f_L
             \fh(1+f_{\widetilde{N}_i}) \fNi \right] \; , 
\end{eqnarray}
\begin{eqnarray}
s \bar{D}_i & = & \Lambda^{12}_{\widetilde{N}_i}
             \left[f_{\widetilde{N}_i}(1-f_{\bar{L}})(1-\fh)\Nifb
             - \right. \nonumber \\ 
             & & \left. - f_{\bar{L}}
             \fh(1+f_{\widetilde{N}_i}) \fbNi \right] \; ,
\end{eqnarray}
\begin{eqnarray}
s \widetilde{D}_i & = &
             \Lambda^{12}_{\widetilde{N}_i}
             \left[f_{\widetilde{N}_i}(1+f_{\widetilde{L}})(1+\fH)\Nis
             - \right. \nonumber \\ 
             & & \left. - f_{\widetilde{L}} \fH
             (1+f_{\widetilde{N}_i})\sNi \right] \; , 
\end{eqnarray}
\begin{eqnarray}
s \widetilde{D}^{\dag}_i & = &
             \Lambda^{12}_{\widetilde{N}_i}
             \left[f_{\widetilde{N}_i}(1+f_{\widetilde{L}^{\dag}})
             (1+\fH)\Nisb- \right. \nonumber \\ 
             & &
             \left. -f_{\widetilde{L}^{\dag}}
             \fH(1+f_{\widetilde{N}_i})\sbNi \right] \; , 
\end{eqnarray}
and
\begin{eqnarray}
s S & = & \Lambda^{12}_{34}
             \left[f_L \fh (1-f_{\bar{L}})(1-\fh)
{\displaystyle \sum_{k,k'}}
|M_{sub}(L_k h
             \to \bar{L}_{k'} \bar{h})|^2- \right. \nonumber \\ 
             & & \left.-
             f_{\bar{L}} \fh (1-f_L)(1-\fh)
{\displaystyle \sum_{k,k'}} |M_{sub}(\bar{L}_k
             \bar{h} \to L_{k'} h)|^2 \right] \; ,   
\label{eq:S1}
\end{eqnarray}
\begin{eqnarray}
s\widetilde{S} & = & \Lambda^{12}_{34}
             \left[f_{\widetilde{L}}\fH
             (1+f_{\widetilde{L}^{\dag}})(1+\fH)
{\displaystyle \sum_{k,k'}}          
|M_{sub}(\widetilde{L}_k H \to \widetilde{L}_{k'}^{\dag}
             H^{\dag})|^2- \right. \nonumber \\ 
             & & \left.-
             f_{\widetilde{L}^{\dag}} \fH
             (1+f_{\widetilde{L}})(1+\fH)
{\displaystyle \sum_{k,k'}}
|M_{sub}(\widetilde{L}_k^{\dag}
             H^{\dag} \to \widetilde{L}_{k'} H)|^2 \right] \; , 
\label{eq:S2}
\end{eqnarray}
\begin{eqnarray}
s S_{L\widetilde{L}^{\dag}} & = &
             \Lambda^{12}_{34} \left[f_L \fh
             (1+f_{\widetilde{L}^{\dag}})(1+f_{H^{\dag}})|
{\displaystyle \sum_{k,k'}}|M_{sub}(L_k h
             \to \widetilde{L}_{k'}^{\dag} H^{\dag})|^2- \right. \nonumber
             \\ 
             & & \left.- f_{\widetilde{L}^{\dag}}
             \fH(1-f_L)(1-\fh)
{\displaystyle \sum_{k,k'}}|M_{sub}(\widetilde{L}_k^{\dag}
             H^{\dag} \to L_{k'} h)|^2 \right] \; ,   
\label{eq:S3}
\end{eqnarray}
\begin{eqnarray} 
s\bar{S}_{L\widetilde{L}^{\dag}} & = & \Lambda^{12}_{34}
             \left [ f_{\bar{L}} \fh
             (1+f_{\widetilde{L}})(1+\fH) 
{\displaystyle \sum_{k,k'}}|M_{sub}(\bar{L}_k \bar{h} \to
             \widetilde{L}_{k'} H)|^2- \right.\nonumber \\ 
             & & \left. -
             f_{\widetilde{L}} \fH (1-f_{\bar{L}})(1-\fh)
{\displaystyle \sum_{k,k'}} 
|M_{sub}(\widetilde{L}_k H \to \bar{L}_{k'} \bar{h})|^2 \right]
             \; ,  
\label{eq:S4}
\end{eqnarray}
\begin{eqnarray} 
s S_{L\widetilde{L}} & = &
             \Lambda^{12}_{34} \left[f_L \fh
             (1+f_{\widetilde{L}})(1+\fH)
{\displaystyle \sum_{k,k'}}|M_{sub}(L_k h \to
             \widetilde{L}_{k'} H)|^2- \right. \nonumber \\ 
             & & \left.-
             f_{\widetilde{L}} \fH(1-f_L)(1-\fh)
{\displaystyle \sum_{k,k'}}|M_{sub}(\widetilde{L}_k
             H \to L_{k'} h)|^2 \right]  \; ,   
\label{eq:S5}
\end{eqnarray}
\begin{eqnarray} 
s  \bar{S}_{L\widetilde{L}} & = & \Lambda^{12}_{34} \left [
             f_{\bar{L}} \fh
             (1+f_{\widetilde{L}^{\dag}})(1+\fH)
{\displaystyle \sum_{k,k'}} 
|M_{sub}(\bar{L}_k \bar{h} \to \widetilde{L}_{k'}^{\dag}
             H^{\dag})|^2- \right. \nonumber \\ & & \left.-
             f_{\widetilde{L}^{\dag}} \fH
             (1-f_{\bar{L}})(1-\fh)
{\displaystyle \sum_{k,k'}}
|M_{sub}(\widetilde{L}_k^{\dag} H^{\dag} \to \bar{L}_{k'}\bar{h})|^2 \right] \; .
\label{eq:S6}
\end{eqnarray}
The $S-terms$ are defined in terms of subtracted amplitudes,
since the on-shell contribution is already taken into account through
the decays and inverse decays in the $D-terms$. So for example:

\begin{equation}
\left|M_{sub}(L_k h \to \bar{L}_{k'} \bar{h})\right|^2=
\left|M(L_k h \to \bar{L}_{k'} \bar{h})
\right|^2-
\left| M_{os}(L_kh \to \bar{L}_{k'} \bar{h}) \right|^2 ~,
\end{equation}
where,
\begin{equation}
\left| M_{os}(L_kh \to \bar{L}_{k'} \bar{h}) \right|^2 = 
\left| \hat{A} (L_k h \to\widetilde{N}_i)\right|^2 
\frac{\pi \delta(s-m_{\widetilde{N}_i})}{m_{\widetilde{N}_i}  
\Gamma_{\widetilde{N}_i}} 
\left| \hat{A} (\widetilde{N}_i \to \bar{L}_{k'} \bar{h} )\right|^2 
\; .
 \end{equation}  

In writing Eqs.(\ref{eq:bolN})--(\ref{eq:bolLs}) we have not included
the $\Delta L=1$ processes. They do not contribute to  
the out of equilibrium condition. However they can lead to 
a dilution of the generated ${\cal L}_{total}$. Therefore they are relevant
in the exact computation of the $\kappa$ factor defined in 
Eq.~(\ref{eq:yb-l}).

In order to compute the $D-terms$, we use the following relation between 
the equilibrium densities:
\bea 
\label{rel_dis:eq2} 
f_{\overline{L}^{\!\!\!\!\!{(\;\;)}}} 
\fh
 (1+\fNieq) = \fNieq 
(1- f_{\overline{L}^{\!\!\!\!\!{(\;\;)}}})
(1-\fh) e^{\mp \mu_f/T}
 \simeq \nonumber \\ 
\simeq \fNieq 
(1- f^{eq}_L)
(1-\fh)(1 \mp \YLf) \; , \\ 
\nonumber \\
f_{\widetilde{L}^{(\dag)}} \fH 
(1+\fNieq) 
= \fNieq (1 + f_{\widetilde{L}^{(\dag)}})(1+\fH)e^{\mp \mu_s/T}
 \simeq \nonumber \\ 
\simeq \fNieq (1 + f^{eq}_{\widetilde{L}})(1+\fH)
(1 \mp \YLs)
\;, 
\eea
with
\begin{equation}
\fNieq=\frac{1}{\exp[E_{\widetilde{N_i}}/T]-1} \; .
\end{equation}
One gets
\bea
D_i+\bar{D}_i
 & = & \frac{1}{s}\Lambda_{\widetilde{N}_i}^{12} 
\left[ f_{\widetilde{N}_i}(1-f^{eq}_L)(1-\fh) 
\left( \frac{1+\efi}{2}+\frac{1-\efi}{2}  \right)\Af  \right. \nonumber \\
 &  & - \left. \fNieq \frac{1+f_{\widetilde{N}_i}}{1+\fNieq} 
(1-f^{eq}_L)(1-\fh)
\left[ (1-\YLf)\frac{1-\efi}{2}+(1+\YLf)\frac{1+\efi}{2}  \right] \Af \right] 
= \nonumber \\
 & = & 
\left(Y_{\widetilde{N}_i}\langle \Gamma_{\widetilde{N}_i}^{f} \rangle 
-\YNieq \langle \widetilde{\Gamma}_{\widetilde{N}_i}^{f} \rangle
\right) + \YNieq \YLf \efi  
\langle \widetilde{\Gamma}_{\widetilde{N}_i}^{f} \rangle  \nonumber
\eea
where in order to write the equations in the closest to the standard notation 
we have defined the following average widths:
\bea
n_{\widetilde{N}_i}^{eq} \langle \Gamma_{\widetilde{N}^{eq}_i}^{f}\rangle
&=&\Lambda_{\widetilde{N}_i}^{12} \fNieq(1-\fLeq)(1-\fh) \Af  \\
n_{\widetilde{N}_i} \langle \Gamma_{\widetilde{N}_i}^{f}\rangle
&=&\Lambda_{\widetilde{N}_i}^{12} \ffNi (1-\fLeq)(1-\fh) \Af  \\
n_{\widetilde{N}_i}^{eq} \langle 
\widetilde{\Gamma}_{\widetilde{N}_i}^{f}\rangle
&=&\Lambda_{\widetilde{N}_i}^{12} \fNieq \frac{(1+\ffNi)}{(1+\fNieq)}
(1-\fLeq)(1-\fh) \Af  \\
n_{\widetilde{N}_i}^{eq} \langle \Gamma_{\widetilde{N}^{eq}_s}^{s}\rangle
&=&\Lambda_{\widetilde{N}_i}^{12} \fNieq(1+\fLteq)(1+\fH) \As  \\
n_{\widetilde{N}_i} \langle \Gamma_{\widetilde{N}_i}^{s}\rangle
&=&\Lambda_{\widetilde{N}_i}^{12} \ffNi (1+\fLeq)(1+\fH) \As  \\
n_{\widetilde{N}_i}^{eq} \langle 
\widetilde{\Gamma}_{\widetilde{N}_i}^{s}\rangle
&=&\Lambda_{\widetilde{N}_i}^{12} \fNieq \frac{(1+\ffNi)}{(1+\fNieq)}
(1+\fLteq)(1+\fH) \As  
\eea
which verify that in equilibrium 
\begin{equation}
\langle{\Gamma}_{\widetilde{N}_i}^{f(s)}\rangle=
\langle {\Gamma}_{\widetilde{N}_i^{eq}}^{f(s)}\rangle=
\langle\widetilde{\Gamma}_{\widetilde{N}_i}^{f(s)}\rangle
\label{eqgams}
\end{equation}

Equivalently for the rest of terms :
\bea
D_i+\bar{D}_i & = &
\left(Y_{\widetilde{N}_i}\langle \Gamma_{\widetilde{N}_i}^{f} \rangle 
-\YNieq \langle \widetilde{\Gamma}_{\widetilde{N}_i}^{f} \rangle
\right) + \YNieq \YLf \efi  
\langle \widetilde{\Gamma}_{\widetilde{N}_i}^{f} \rangle  \\
D_i-\bar{D}_i & = & 
\efi \left(Y_{\widetilde{N}_i}\langle \Gamma_{\widetilde{N}_i}^{f} \rangle 
+\YNieq \langle \widetilde{\Gamma}_{\widetilde{N}_i}^{f} \rangle
\right) + \YNieq \YLf 
\langle \widetilde{\Gamma}_{\widetilde{N}_i}^{f} \rangle  \\
\widetilde{D}_i+\widetilde{D}_i^{\dag}  & = & 
\left(Y_{\widetilde{N}_i}\langle \Gamma_{\widetilde{N}_i}^{s} \rangle 
-\YNieq \langle \widetilde{\Gamma}_{\widetilde{N}_i}^{s} \rangle
\right) + \YNieq \YLs \esi  
\langle \widetilde{\Gamma}_{\widetilde{N}_i}^{s} \rangle  \\
\widetilde{D}_i-\widetilde{D}_i^{\dag}  & = & 
\esi \left(Y_{\widetilde{N}_i}\langle \Gamma_{\widetilde{N}_i}^{s} \rangle 
+\YNieq \langle \widetilde{\Gamma}_{\widetilde{N}_i}^{s} \rangle
\right) + \YNieq \YLs   
\langle \widetilde{\Gamma}_{\widetilde{N}_i}^{s} \rangle  
\eea

Concerning $S-terms$, 
in order to  evaluate, for example, 
the on-shell contribution $\left| M_{os}(L_kh \to
\bar{L}_{k'} \bar{h}) \right|^2$ we use the following relation between the
equilibrium densities 
\be \label{rel_dis:eq1} 
(1-\fLeq)(1-\fh) = \fNieq e^{E_N/T}\left[(1-\fLeq)(1-\fh)-
\fLeq \fh \right] ~,
\ee
and the identity,
\be
1 = \int d^4p_N \delta^4 (p_{\widetilde{N}_i}-p_{{L}}-
p_{{h}}) ~.
\ee
They allow us to write  the on-shell contribution to 
the scattering terms at the required order in $\epsilon$ as:
\bea
\Lambda_{12}^{34}~ f_L \fh (1-f_{\bar{L}})(1-\fh) \fNi
\frac{\pi \delta(s-m_{\widetilde{N}_i})}{m_{\widetilde{N}_i}
\Gamma^{th}_{\widetilde{N}_i}} \Nifbp  = \nonumber \\
 = \int \frac{d^3 p_L}{(2\pi)^3 2E_L} \frac{d^3 p_h}{(2\pi)^3 2E_h}(2
 \pi)^4 \delta^4 (p_{\widetilde{N}_i}-p_L-p_h ) f^{av}_L \fh
 \left( \frac{1-\efi}{2}\right)^2 \Af \times \nonumber \\
 \int \frac{d^4 p_{\widetilde{N}_i}}{(2 \pi)^4} \frac{2 \pi
 \delta(s-m_{\widetilde{N}_i})}{2 m_{\widetilde{N}_i}
 \Gamma^{th}_{\widetilde{N}_i}} \int \frac{d^3 p_{{L}}}{(2\pi)^3
 2E_{{L}}} \frac{d^3 p_{{h}}}{(2\pi)^3 2E_{{h}}}(2 \pi)^4
 \delta^4 (p_{\widetilde{N}_i}-p_{{L}}-p_{{h}} )
 \times \nonumber \\
\fNieq e^{E_N/T}\left[(1-\fLeq)(1-\fh)-\fLeq \fh \right] \ \Afp 
~~~~~~~~~~~~~~~~~~~~~~~~~  \; .
\label{eq:onshell}
\eea 
Now we define the thermal width into fermion and scalars as:
\bea
\Gamma^{th,f}_{\widetilde{N}_i} & = & \frac{1}{2
m_{\widetilde{N}_i}}\int \frac{d^3p_{{L}}}{(2 \pi)^3 2
E_{{L}}} \frac{d^3p_{{h}}}{(2 \pi)^3 2 E_{{L}}}(2 \pi)^4
\delta^4(p_{\widetilde{N}_i}-p_{{L}}-p_{{h}})
\nonumber \\ 
&& 
\left[(1-f_{{L}}^{eq})(1-\fh)-f_{{L}}^{eq}  \fh
\right] \Af \; ,\nonumber \\ 
\Gamma^{th,s}_{\widetilde{N}_i} & = &
\frac{1}{2 m_{\widetilde{N}_i}}\int \frac{d^3p_{\widetilde{L}}}{(2
\pi)^3 2 E_{\widetilde{L}}} \frac{d^3p_H}{(2 \pi)^3 2 E_H}(2 \pi)^4
\delta^4(p_{\widetilde{N}_i}-p_{\widetilde{L}}-p_H)
\nonumber \\ 
&&
\left[(1+f_{\widetilde{L}}^{eq})(1+\fH) - f_{\widetilde{L}}^{eq}\fH
\right] \As \; .
\eea
so that the thermal width of sneutrinos is
$\Gamma^{th}_{\widetilde{N}_i}=\Gamma^{th,f}_{\widetilde{N}_i}
+\Gamma^{th,s}_{\widetilde{N}_i}$. 

Using 
that $\int \frac{d^4 p_{\widetilde{N}_i}}{(2 \pi)^4} 2\pi
\delta (p_{\widetilde{N}_i}^2
-m_{\widetilde{N}_i}^2 )=\int\frac{d^3p_{\widetilde{N}_i}}{(2
\pi)^3 2 E_{\widetilde{N}_i}}$ we can write Eq.~(\ref{eq:onshell}) 
as:
\bea
\Lambda_{\widetilde{N}_i}^{12} ~ \fNieq(1-f_L^{eq})(1-\fh)\left(
\frac{1-\efi}{2}\right)^2 \Af
\frac{\Gamma^{th,f}_{\widetilde{N}_i}}{\Gamma^{th}_{\widetilde{N}_i}}=
n_{\widetilde{N}_i}^{eq}  \frac{(1-\efi)^2}{4}
\langle \Gamma_{\widetilde{N}_i^{eq}}^{f}\rangle
\frac{\Gamma^{th,f}_{\widetilde{N}_i}}{\Gamma^{th}_{\widetilde{N}_i}}
~. \nonumber
\eea
Altogether we find that 
\begin{eqnarray}
S & = & 
{\displaystyle \sum_{k,k'}}
\langle \sigma(L_k h \to \bar{L}_{k'} \bar{h})-\sigma(\bar{L}_{k'} 
\bar{h} \to L_k h)\rangle  
+\YNieq \efi \langle \Gamma_{\widetilde{N}_i^{eq}}^{f} \rangle 
\frac{\Gamma^{th,f}_{\widetilde{N}_i}}{\Gamma^{th}_{\widetilde{N}_i}} \; .  
\end{eqnarray}
The rest of on-shell contributions can be evaluated similarly: 
\begin{eqnarray}
\widetilde{S} & = & 
{\displaystyle \sum_{k,k'}}
\langle \sigma(\widetilde{L}_k H \to \widetilde{L}_{k'}^{\dag} 
H^{\dag})-\sigma(\widetilde{L}_{k'}^{\dag} H^{\dag} \to \widetilde{L}_k H)
\rangle 
+\YNieq \esi \langle \Gamma_{\widetilde{N}_i^{eq}}^{s} \rangle 
\frac{\Gamma^{th,s}_{\widetilde{N}_i}}{\Gamma^{th}_{\widetilde{N}_i}}    
\end{eqnarray}
\begin{eqnarray}
S_{L\widetilde{L}^{\dag}} & = & 
{\displaystyle \sum_{k,k'}}
\langle \sigma(L_k h \to \widetilde{L}_{k'}^{\dag} H^{\dag})
-\sigma(\widetilde{L}_{k'}^{\dag} H^{\dag} \to L_k h)\rangle 
+\YNieq \frac{\efi +\esi}{2} \langle \Gamma^f_{\widetilde{N}_i^{eq}}\rangle 
\frac{\Gamma^{th,s}_{\widetilde{N}_i}}{\Gamma^{th}_{\widetilde{N}_i}} 
= \nonumber \\
  & = &  
{\displaystyle \sum_{k,k'}}
\langle \sigma(L_k h \to \widetilde{L}_{k'}^{\dag} H^{\dag})
-\sigma(\widetilde{L}_{k'}^{\dag} H^{\dag} \to L_k h)\rangle 
+\YNieq \frac{\efi +\esi}{2}
\langle \Gamma_{\widetilde{N}_i^{eq}}^{s} \rangle 
\frac{\Gamma^{th,f}_{\widetilde{N}_i}}{\Gamma^{th}_{\widetilde{N}_i}}  
\end{eqnarray}
\begin{eqnarray}
\bar{S}_{L\widetilde{L}^{\dag}} & = &
{\displaystyle \sum_{k,k'}} 
\langle \sigma(\bar{L}_k \bar{h} \to \widetilde{L}_{k'} H)
-\sigma(\widetilde{L}_{k'} H \to \bar{L}_k \bar{h})\rangle 
- \YNieq \frac{\efi +\esi}{2} 
\langle \Gamma^f_{\widetilde{N}_i^{eq}} \rangle 
\frac{\Gamma^{th,s}_{\widetilde{N}_i}}{\Gamma^{th}_{\widetilde{N}_i}} 
= \nonumber \\
& = & 
{\displaystyle \sum_{k,k'}}
\langle \sigma(\bar{L}_k \bar{h} \to \widetilde{L}_{k'} H)-
\sigma(\widetilde{L}_{k'} H \to \bar{L}_k \bar{h})\rangle  
- \YNieq \frac{\efi +\esi}{2} \langle \Gamma_{\widetilde{N}_i^{eq}}^{s} 
\rangle 
\frac{\Gamma^{th,f}_{\widetilde{N}_i}}{\Gamma^{th}_{\widetilde{N}_i}}    
\end{eqnarray}
\begin{eqnarray}
S_{L\widetilde{L}} & = & 
{\displaystyle \sum_{k,k'}}
\langle\sigma(L_k h \to \widetilde{L}_{k'} H)
-\sigma(\widetilde{L}_{k'} H \to L_k h)\rangle + \YNieq \frac{\efi - \esi}{2} 
\langle \Gamma^f_{\widetilde{N}_i^{eq}}\rangle 
\frac{\Gamma^{th,s}_{\widetilde{N}_i}}{\Gamma^{th}_{\widetilde{N}_i}} = 
\nonumber \\
& = & 
{\displaystyle \sum_{k,k'}} 
\langle \sigma(L_k h \to \widetilde{L}_{k'} H)
-\sigma(\widetilde{L}_{k'} H \to L_k h)\rangle  - \YNieq \frac{\efi - \esi}{2} 
\langle \Gamma_{\widetilde{N}_i^{eq}}^{s}\rangle  
\frac{\Gamma^{th,f}_{\widetilde{N}_i}}
{\Gamma^{th}_{\widetilde{N}_i}}  
\end{eqnarray}
\begin{eqnarray}
\bar{S}_{L\widetilde{L}} & = & 
{\displaystyle \sum_{k,k'}}
\langle \sigma(\bar{L}_k \bar{h} \to \widetilde{L}_{k'}^{\dag} H^{\dag})
-\sigma(\widetilde{L}_{k'}^{\dag} H^{\dag} \to \bar{L}_k \bar{h})\rangle - 
\YNieq \frac{\efi - \esi}{2}\langle \Gamma^f_{\widetilde{N}_i^{eq}}\rangle 
\frac{\Gamma^{th,s}_{\widetilde{N}_i}}{\Gamma^{th}_{\widetilde{N}_i}} 
= \nonumber \\
  & = & 
{\displaystyle \sum_{k,k'}}
\langle\sigma(\bar{L}_k \bar{h} \to \widetilde{L}_{k'}^{\dag} H^{\dag})
-\sigma(\widetilde{L}_{k'}^{\dag} H^{\dag} \to \bar{L}_k \bar{h})\rangle   
- \YNieq \frac{\efi - \esi}{2} \langle \Gamma_{\widetilde{N}_i^{eq}}^{s}\rangle 
\frac{\Gamma^{th,f}_{\widetilde{N}_i}}{\Gamma^{th}_{\widetilde{N}_i}}   
\end{eqnarray}

Altogether we can write the Boltzmann equations for the sneutrinos 
and leptonic numbers
\footnote {Here we have suppressed flavour indices in the two body 
$\Delta L =2$ scattering terms for
the sake of simplicity, but a sum over all flavours in initial and
final states should be understood.}:
\bea
\frac{d Y_{\widetilde{N}_i}}{d t} & = & - Y_{\widetilde{N}_i}
\left(  \langle\Gamma_{\widetilde{N}_i}^{f}\rangle + 
\langle \Gamma_{\widetilde{N}_i}^{s}\rangle \right) 
+\YNieq 
\left(  \langle\widetilde{\Gamma}_{\widetilde{N}_i}^{f}\rangle + 
\langle \widetilde{\Gamma}_{\widetilde{N}_i}^{s}\rangle \right) 
\nonumber\\ 
&&
-\YNieq \left(
\YLf \efi  \langle \widetilde{\Gamma}_{\widetilde{N}_i}^{f}\rangle + 
\YLs \esi  \langle \widetilde{\Gamma}_{\widetilde{N}_i}^{s}\rangle \right) 
\label{eqyxfin}
\end{eqnarray}
\begin{eqnarray}
\frac{d \YLf}{d t} & = & \sum_i \left[
\efi 
\left(Y_{\widetilde{N}_i}\langle \Gamma_{\widetilde{N}_i}^{f} \rangle 
+\YNieq \langle \widetilde{\Gamma}_{\widetilde{N}_i}^{f} \rangle
-2  \YNieq \langle \Gamma_{\widetilde{N}^{eq}_i}^{f}
\rangle
\right) + \YNieq \YLf 
\langle \widetilde{\Gamma}_{\widetilde{N}_i}^{f} \rangle  \right]
\nonumber \\
&& -  \langle 2\sigma(L h \to \bar{L} \bar{h}) +
\sigma(L h \to \widetilde{L}^{\dag} H^{\dag})+ 
\sigma(L h \to \widetilde{L} H) \rangle  \nonumber \\
 & &+  \langle 2 \sigma(\bar{L} \bar{h} \to L h) 
+ \sigma(\bar{L} \bar{h} \to \widetilde{L} H) 
+ \sigma(\bar{L} \bar{h} \to \widetilde{L}^{\dag} H^{\dag})\rangle 
 \nonumber \\
 && -  \langle\sigma(\widetilde{L} H \to \bar{L} \bar{h}) 
- \sigma(\widetilde{L} H \to L h) \rangle + \nonumber \\
 && +  \langle\sigma(\widetilde{L}^{\dag} H^{\dag} \to L h) 
-  \sigma(\widetilde{L}^{\dag} H^{\dag} \to \bar{L} \bar{h})\rangle  
\label{eqylffin}
\end{eqnarray}
\begin{eqnarray}
\frac{d \YLs}{d t} & = & \sum_i 
\left[ 
\esi \left(Y_{\widetilde{N}_i}\langle \Gamma_{\widetilde{N}_i}^{s} \rangle 
+\YNieq \langle \widetilde{\Gamma}_{\widetilde{N}_i}^{s} \rangle
-2  \YNieq \langle \Gamma_{\widetilde{N}^{eq}_i}^{s}
\rangle
\right) + \YNieq \YLs 
\langle \widetilde{\Gamma}_{\widetilde{N}_i}^{s} \rangle  \right]
\nonumber \\
&& -  \langle2\sigma(\widetilde{L} H \to \widetilde{L}^{\dag} H^{\dag})
+\sigma(\widetilde{L} H \to \bar{L} \bar{h})+ 
\sigma(\widetilde{L} H \to L h) \rangle  \nonumber \\
&& +\langle 2\sigma(\widetilde{L}^{\dag} H^{\dag} \to 
\widetilde{L} H)+\sigma(\widetilde{L}^{\dag} H^{\dag} \to L h)
+\sigma(\widetilde{L}^{\dag} H^{\dag} \to \bar{L} \bar{h}) \rangle \nonumber \\
&& -  \langle\sigma(Lh \to \widetilde{L}^{\dag} H^{\dag})
-\sigma(L h \to \widetilde{L} H) \rangle  \nonumber \\
&& + \langle \sigma(\bar{L} \bar{h} \to \widetilde{L} H) 
-  \sigma(\bar{L} \bar{h} \to \widetilde{L}^{\dag} H^{\dag}) \rangle  
\label{eqylsfin}
\eea
The out of equilibrium condition is verified since using Eq.~(\ref{eqgams}) 
the first term of Eq.~(\ref{eqyxfin}) and the $\epsilon$ terms of
Eqs.(\ref{eqylffin}) and (\ref{eqylsfin}) cancel out in thermal equilibrium.

The Boltzmann equation for the total lepton number can be written as 
(here we use $\esi=-\efi=\overline{\epsilon}_i$):
\bea
\frac{d Y_{{\cal L}_{total}}}{d t} & = & 
\sum_i \overline{\epsilon}_i \left[
Y_{\widetilde{N}_i}\left(
\langle \Gamma_{\widetilde{N}_i}^{s} \rangle
- \langle \Gamma_{\widetilde{N}_i}^{f} \rangle\right)
+\YNieq \left(\langle \widetilde{\Gamma}_{\widetilde{N}_i}^{s} \rangle
-\langle \widetilde{\Gamma}_{\widetilde{N}_i}^{f} \rangle\right)
-2  \YNieq 
\left(\langle \Gamma_{\widetilde{N}^{eq}_i}^{s}\rangle
-\langle \Gamma_{\widetilde{N}^{eq}_i}^{f}\rangle \right)
\right.
\nonumber \\
  && + \left.   \YNieq(\YLf \widetilde{\Gamma}_{\widetilde{N}_i}^{f}+\YLs 
\widetilde{\Gamma}_{\widetilde{N}_i}^{s}) 
\right]  +  ~ scattering ~  terms \nonumber \\
&\simeq&
\sum_i \left[\langle \Gamma_{\widetilde{N}^{eq}_i} \rangle
\left(Y_{\widetilde{N}_i}-\YNieq \right) \;\epsilon^{eff}_i(T)
+ \YNieq \left( \YLf \widetilde{\Gamma}_{\widetilde{N}^{eq}_i}^{f}+\YLs 
\widetilde{\Gamma}_{\widetilde{N}_i}^{s} \right) 
\right]  +  ~ s. ~  t.
\label{eq:ltotal}
\eea
with $\langle \Gamma_{\widetilde{N}^{eq}_i}\rangle=
\langle \Gamma_{\widetilde{N}^{eq}_i}^{f}\rangle
+\langle \Gamma_{\widetilde{N}^{eq}_i}^{s}\rangle$.

In the last line we have used that at 
${\cal O}(\epsilon)$ we can neglect the difference 
between  $\ffNi$ and $\fNieq$ in the definitions of the thermal
average widths 
and we have defined the effective $T$ dependent
total asymmetry:
\begin{equation}
\epsilon^{eff}_i (T)=\overline{\epsilon}_i \frac
{\langle \Gamma_{\widetilde{N}_i}^{s}\rangle
-\langle \Gamma_{\widetilde{N}^{eq}_i}^{f}\rangle}
{\langle \Gamma_{\widetilde{N}^{eq}_i}^{f}\rangle
+\langle \Gamma_{\widetilde{N}^{eq}_i}^{s}\rangle}\;,
\label{epsieff}
\end{equation}
which in the approximate decay at rest takes the form
Ref.\cite{soft1,soft2}
\begin{equation}
\epsilon_i(T)=\overline{\epsilon}_i  \frac{c_s-c_f}{c_s+c_f}\; .
\end{equation}

\section{Results}
\label{res}

We now quantify the conditions on the parameters which can be
responsible for a successful leptogenesis.

The final amount of ${\cal B}-{\cal L}$ asymmetry $Y_{{\cal B}-{\cal
L}}=n_{{\cal B}-{\cal L}}/s$ generated by the decay of the four light
singlet sneutrino states $\tilde N_i$ assuming no pre-existing
asymmetry and thermal initial sneutrino densities 
can be parameterized as
\begin{equation}
Y_{{\cal B}-{\cal L}}=-\kappa\, \sum_i \, \epi(T_d)\, 
Y^{\rm eq}_{\tilde N_i} (T\gg M_i) \, .
\label{eq:yb-l}
\end{equation}
$\epi(T)$ is given in Eq.(\ref{asymi}) and $T_d$ is the temperature
at the time of decay defined by the condition that the decay 
width is equal to the expansion rate of the universe: $\Gamma=H(T_d)$, 
where the Hubble parameter $H=1.66~g_{*}^{1/2} \frac{T^2}{m_{pl}}$, 
$m_{pl}=1.22\cdot 10^{19}~$GeV is the Planck mass and 
$g_*$ counts the effective number of
spin-degrees of freedom in thermal equilibrium, $g_{*}=228.75$ in the MSSM.
Furthermore $Y^{\rm eq}_{\tilde N_i}(T\gg M_i) = 90 \zeta(3)/(4\pi^4g_*)$.

In Eq.~(\ref{eq:yb-l}) $\kappa\lesssim 1$ is a dilution factor which
takes into account the possible inefficiency in the production of the
singlet sneutrinos, the erasure of the generated asymmetry by
$L$-violating scattering processes and the temperature dependence of the 
CP asymmetry $\epsilon_i(T)$. 
The precise value of $\kappa$ can only be obtained from numerical
solution of the Boltzmann equations. Moreover, in general, the
result depends on how the lepton asymmetry is distributed in the three
lepton flavours~\cite{flavour}.  For simplicity we will ignore flavour
issues. Furthermore, in what follows we will use an approximate 
constant value $\kappa=0.2$.

After conversion by the sphaleron transitions, the final baryon asymmetry
is related to the ${\cal B}-{\cal L}$ asymmetry by
\begin{equation}
\frac{n_{\cal B}}{s}=\frac{24+4n_H}{66+13n_H}\frac{n_{{\cal B} - {\cal
L}}}{s},
\end{equation}
where $n_H$ is the number of Higgs doublets. For the MSSM:
\begin{equation}
\frac{n_{\cal B}}{s} =-8.4\times 10^{-4}\,\kappa \, \sum_i \epsilon_i(T_d)
\end{equation}

This has to be compared with the WMAP measurements that in
the $\Lambda$CDM model imply~\cite{WMAP}: 
\begin{equation}
\frac{n_{\cal B}}{s} = (8.7^{+0.3}_{-0.4})\times 10^{-11}\qquad
\label{eq:b/sexp}
\end{equation}

Altogether we find that (for maximal CP violating phase $\sin\phi=1$):
\begin{equation}
\frac{4\, |B_{SN}\, A|\, \Gamma }{4 |B_{SN}|^2 + |M|^2 \Gamma^2} \,
\frac{c_s(T_d) - c_f(T_d)}{c_s(T_d) + c_f(T_d)} \gtrsim 2.6\times 10^{-7} 
\label{eq:epsibound}
\end{equation}

Further constraints arise from the timing of
the decay. First, successful leptogenesis requires the singlet 
sneutrinos to decay out of 
equilibrium: its decay width must be smaller than the expansion rate 
of the Universe
$\Gamma < H\left|_{T=M} \right.$, with $\Gamma$ given in Eq.~(\ref{eq:gamma}),
\begin{equation}
\frac{M \displaystyle \sum_k 
|Y_{1k}|^2}{8\pi}<1.66~ g_{*}^{1/2}\frac{M^2}{m_{pl}}\; .
\label{eq:outeq}
\end{equation}
This condition gives an upper bound:
\begin{equation} 
\label{cotainf}
\sum_k |Y_{1k}|^2\, \left(\frac{10^{8}\,{\rm GeV}}{M}\right) \, 
<5\times 10^{-9}
\end{equation}

Second, in order for the generated lepton asymmetry to be 
converted into a baryon asymmetry via the B-L violating sphaleron
processes, the singlet sneutrino decay should occur before the electroweak 
phase transition  
\begin{equation}
\Gamma > H(T \sim 100~{\rm GeV})\; \Rightarrow\; 
M \sum_k |Y_{1k}|^2 \geq 2.6 \times 10^{-13} \; {\rm GeV}
\label{Ymax} 
\end{equation}

The combination of Eqs.(\ref{cotainf}) and~(\ref{Ymax})
determines a range for the possible values of $\sum |Y_{1k}|^2$ 
for a given $M$: 
\be 
2.6\times 10^{-21} \; \left( \frac{10^{8}\,{\rm GeV}}{M}\right)
\, < \sum_k |Y_{1k}|^2 \,
< 5 \times 10^{-9}\, \left( \frac{M}{10^{8}\,{\rm GeV}}\right)
\label{eq:ybound}
\ee

We now turn to the consequences that these constraints  
may have for the neutrino mass predictions in this scenario. 
Without loss of generality one can work in the basis in which 
$M_{ij}$ is diagonal. In that basis the light neutrino masses, 
Eq.~(\ref{mnu}),  are:
\begin{equation}
m_{\nu ij} = 3 \times 10^{-3} \, {\rm eV}\, 
\left(\frac{v}{175\,{\rm GeV}}\right)^2
\sum_{kl} Y_{li}  
\frac{10^{8}\,{\rm GeV}}{M_l} \frac{10^{8}\,{\rm GeV}}{M_k} 
\frac{\mu_{kl}}{\rm GeV} Y_{kj}
\label{eq:masa_n}
\end{equation}
where $v=\langle H \rangle$ is the Higgs vev. 

It is clear from Eq.~(\ref{eq:masa_n})  
that the out of equilibrium condition, Eq.~(\ref{cotainf}), 
implies that the contribution of the lightest pseudo-Dirac 
singlet neutrino generation to the neutrino mass is negligible.
Consequently, to reproduce the observed mass differences
$\Delta m^2_\odot$ and $\Delta m^2_{atm}$, the dominant 
contribution to the neutrino masses must arise
from the exchange of the heavier singlet neutrino states.  

This can be easily achieved, for example, in the single right-handed
neutrino dominance mechanism (SRHND) \cite{king}.  These models naturally
explain the strong hierarchy in the masses and the large mixing angle
present in the light neutrino sector. In particular, in the simple
case in which the matrix $\mu$ and $M$ are simultaneously
diagonalizable, the results in Ref.~\cite{king} imply that for the
inverse see-saw model with three generations of singlet neutrinos, the
SRHND condition is attained if there is a strong hierarchy:
\begin{equation}\label{text_diag}
\mu_3 \frac{Y_{3k} Y_{3k'}}{M_3} \gg \mu_2 \frac{Y_{2l} Y_{2l'}}{M_2}
\gg \mu_1 \frac{Y_{1l} Y_{1l'}}{M_1}\; .
\end{equation}
Generically this means that the out of equilibrium
condition requires the neutrino mass spectrum to be strongly hierarchical,
$m_1\ll m_2< m_3$. 

Conversely this implies that the measured neutrino masses do not 
impose any constraint on the combination of Yukawa couplings
and sneutrino masses which is relevant for the generation of the 
lepton asymmetry which can be taken as an independent parameter
in the evaluation of the asymmetry.

Finally we plot in  Fig. \ref{fig:contornos} the range of parameters
$\sum |Y_{1k}|^2$ and $B_{SN}$ for which enough asymmetry is
generated, Eq.~(\ref{eq:epsibound}), and the out of equilibrium
and pre-electroweak phase transition decay  conditions, 
Eq.~(\ref{eq:ybound}) are verified. We show the ranges for 
three  values of $M$  and for the characteristic value of 
$A=m_{SUSY}=10^3$ GeV.
\FIGURE[t]{
\includegraphics[width=0.6\textwidth]{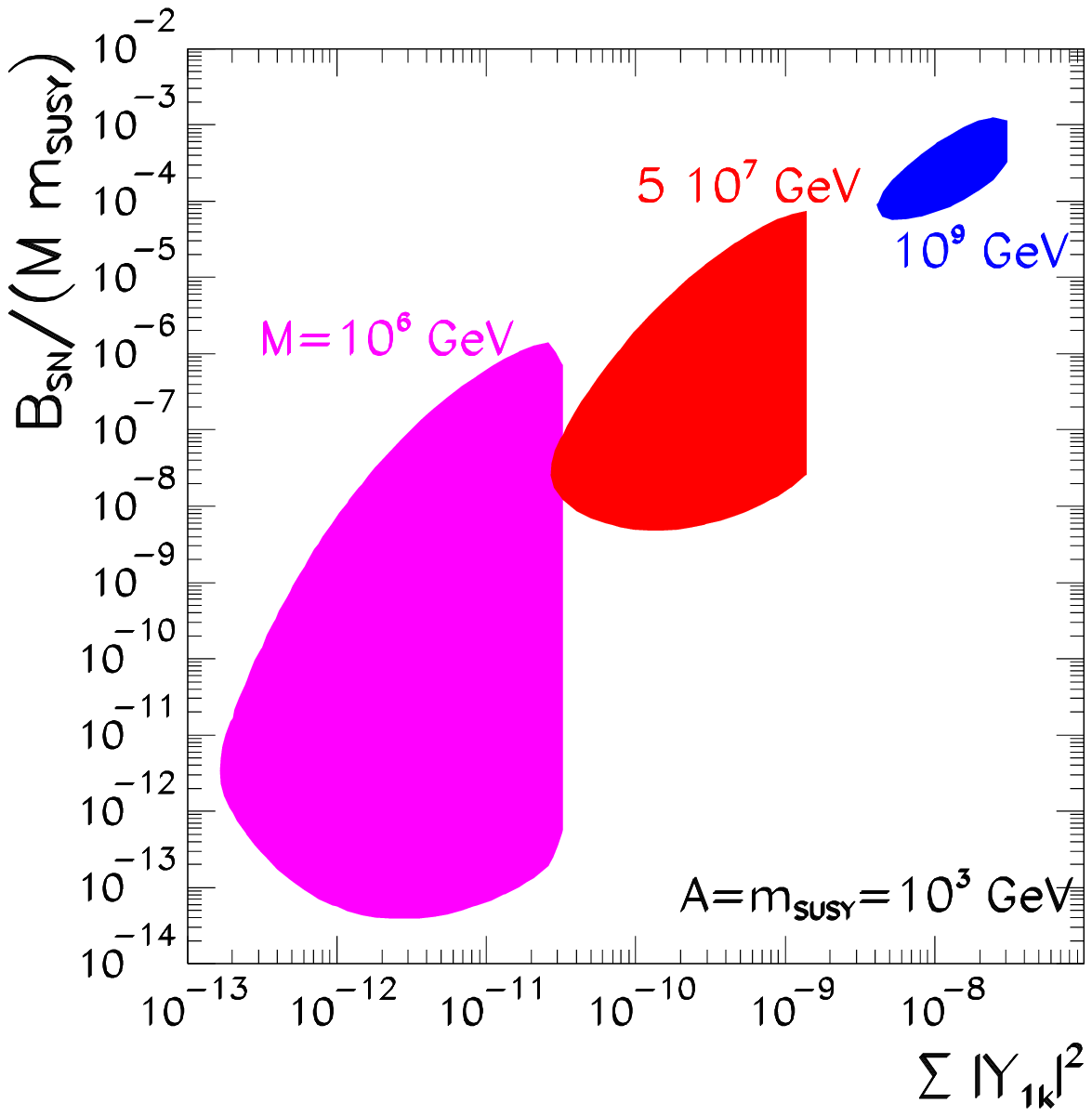}
\caption{${\displaystyle \sum_{k}} |Y_{1k}|^2-B_{SN}$  
regions in which enough $CP$ asymmetry can be generated 
(Eq.~(\ref{eq:epsibound}))
and the non-equilibrium
condition in the sneutrino decay, Eq.~(\ref{eq:outeq}) and decay before
the electroweak phase transition, Eq.~(\ref{Ymax}) 
are verified. We take $A=m_{SUSY}=10^3$ GeV.
The regions correspond to   
$M=10^{6},~ 5\times 10^{7}$, and $10^{9}~{\rm GeV}$, from left to right.
\label{fig:contornos}}}

From the figure we see that this mechanism works for relatively small
values of $M$ ($<10^9~$GeV).  The smaller is $M$, the smaller are the
yukawas $\sum |Y_{1k}|^2$.  Also, in total analogy with the standard
seesaw \cite{soft1}, \cite{soft2}, the value of the soft
supersymmetry-breaking bilinear $B_{SN}$, is well bellow the expected
value $M m_{SUSY}$. The reason is that, in order to generate an
asymmetry large enough $B_{SN} \sim M \Gamma$, but $\Gamma$ is very
small if the sneutrinos decay out of equilibrium, $\Gamma \leqslant 1
~ {\rm GeV} ~ \left( \frac{M}{10^9 ~ {\rm GeV}} \right)^2$. A small 
$B$ term with large CP phase can be  realized naturally for example 
within  the framework of gauge mediated supersymmetry 
breaking~\cite{grossman} or in warped extra dimensions~\cite{wagner}.

Given the small required values of $B_{SN}$ one can question the
expansion in the small parameters in Eq.~(\ref{expanp}).  As described
at the end of Sec.~\ref{qft} in order to verify the stability of the
results we have redone the computation of the CP asymmetry keeping all
the entries in the sneutrino mass matrix, and just assuming that it is
real.  We have found that as long as $|B_{SN}|\gg |B_S|,|\mu|^2$ the
total CP asymmetry is always proportional to $B_{SN}$, and presents
the same resonant behaviour, so that it is still significant only for
$B_{SN} \ll M m_{SUSY}$.

In summary in this work we have studied the conditions for successful
soft leptogenesis  in the context of the supersymmetric inverse seesaw
mechanism. In this model the lepton sector is extended with 
two electroweak singlet superfields to which opposite  
lepton number can be assigned.  This scheme is characterized by a
small lepton number violating Majorana mass term $\mu$ 
with the effective light neutrino mass being $m_\nu \propto \mu$.
The scalar sector contains four single sneutrino states per generation
and, after supersymmetry breaking,  their interaction lagrangian 
contains both $L$-conserving and $L$-violating
soft supersymmetry-breaking bilinear $B$-terms which together with the  $\mu$
parameter give a small mass
splitting between the four singlet sneutrino states of a single
generation. In combination with the trilinear soft supersymmetry 
breaking terms they also provide new CP violating phases needed to
generate a lepton asymmetry in the singlet sneutrino decays.  

We have computed the relevant lepton asymmetry and we find in that, as
long as the $L$-conserving $B$-term, $B_{SN}$, is not much smaller
than the $L$-violating couplings, the asymmetry is proportional to
$B_{SN}$ and it is not suppressed by any $L$-violating parameter.  As
in the standard see-saw case, the asymmetry displays a resonance
behaviour with the maximum value of the asymmetry being obtained when
the largest mass splitting, $2 B_{SN}/M$, is of the order of the
singlet sneutrinos decay width, $\Gamma$. Consequently we find that
this mechanism can lead to successful leptogenesis only for relatively
small values of $B_{SN}$. The right-handed neutrino masses are low
enough to elude the gravitino problem. Also, the out of equilibrium
decay condition implies that the Yukawa couplings involving the
lightest of the right-handed neutrinos are constrained to be very
small which, for the naturally small values of the $L$-violating
parameter $\mu$, implies that the neutrino mass spectrum has to be
strongly hierarchical.

\section*{Acknowledgments}
We would like to thank S. Davidson, M. Hirsch, E. Nardi, Y. Nir, E. Roulet 
and O. Vives for many discussions.  
M.C. G-G and N.R. thank the CERN theory division 
for their hospitality.
J.G. is supported by a MEC-FPU Spanish grant.
This work is supported in part by National Science Foundation
grant PHY-0354776, by the EC RTN network MRTN-CT-2004-503369, 
by Spanish grants 
FPA-2004-00996 and  FPA2005-01269 and by Generalitat Valenciana under grants
ACOMP06/098 and GV05/264.

\newpage

\appendix

\section{CP Asymmetry in Quantum Mechanics }
\label{qm}

The four sneutrino system is completely analogous to the $K^0 -\bar{K}^0$
system, so here we compute the CP asymmetry generated in 
their decay using the same formalism. In order to compare with the 
effective field theory approach described in Sec.~\ref{qft}
we consider only the simplified case  
$B_S, \widetilde{m}_N^2, \widetilde{m}_S^2, \widetilde{m}_{SN}^2  \ll B_{SN}$
and $\widetilde{M}_{SN}^2 \sim \mu M^*$.  
In this limit, we have chosen for simplicity a basis where 
$A=|A| e^{i \phi}$ is the only complex parameter with $\phi$  
given in Eq.(\ref{CP1}).

The evolution of the system is then determined by the effective Hamiltonian,
\be  
\label{H}
H=\hat M-i\frac{\hat\Gamma}{2}
\ee
where, in the interaction basis and at leading order in the expansion 
parameters
$\epsilon,\tilde\epsilon$ defined in Eq.~(\ref{expanp}),
\be
\hat M = M \left(
\begin{array}{cccc}
1&0&\epsilon&\tilde\epsilon \\
0&1&\tilde\epsilon&\epsilon \\
\epsilon&\tilde\epsilon&1& 0 \\
\tilde\epsilon&\epsilon &0&1 
\end{array} \right)
\ee
and 
\be
\label{width}
\hat\Gamma = \Gamma
\left( \begin{array}{cccc} 
1  &  0  &  0  &  \frac{A}{M} \\
0  &  1  &  \frac{A^{*}}{M}  &  0  \\
0  &  \frac{A}{M}  &  1  &  0  \\
\frac{A^{*}}{M}  &  0  &  0  &  1   
\end{array} \right) 
\ee 
with $\Gamma$ given in Eq.(\ref{eq:gamma}).

It is convenient to write the effective Hamiltonian in the mass eigenstate 
basis Eq.(\ref{masseigenstates}), because in such basis the four sneutrino
system decouples in two subsystems of two sneutrinos, with the 
resulting width matrix:
\be   
\label{H_Gamma}
\Gamma = \Gamma
\left(   \begin{array}{cccc}
1-\epsilon_A\cos \phi  &  \epsilon_A\sin \phi   &  0  &  0  \\
\epsilon_A\sin \phi    &  1+\epsilon_A\cos \phi    &  0  &  0  \\
0  &  0  &  1-\epsilon_A\cos \phi  &  - \epsilon_A\sin \phi  \\
0  &  0  &  -\epsilon_A\sin \phi   &  1+\epsilon_A\cos \phi   \\
\end{array}  \right)  \ee
where $\epsilon_A=\frac{\left |A\right |}{|M|}$.

The eigenvectors of the effective Hamiltonian $H$ are:
\begin{equation} 
\begin{array}{ccc}
\widetilde{N}^\prime_1 
& = & \frac{1}{\sqrt{2+2 \frac{|\epsilon_{+}|}{|\epsilon_{-}|}}}
\left[e^{i(\phi_{-}-\phi_{+})/4}(\widetilde{S}^{\dag}-\widetilde{N}^{\dag}) 
+ \sqrt{\frac{|\epsilon_{+}|}{|\epsilon_{-}|}} e^{-i(\phi_{-}-\phi_ {+})/4}
(\widetilde{S}-\widetilde{N}) \right] \nonumber \\ 
\widetilde{N}^\prime_2 
& = & \frac{i}{\sqrt{2+2 \frac{|\epsilon_{+}|}{|\epsilon_{-}|}}}
\left[e^{i(\phi_{-}-\phi_{+})/4}(\widetilde{S}^{\dag}-\widetilde{N}^{\dag}) - 
\sqrt{\frac{|\epsilon_{+}|}{|\epsilon_{-}|}} e^{-i(\phi_{-}-\phi_ {+})/4}
(\widetilde{S}-\widetilde{N}) \right] \nonumber \\ 
\widetilde{N}^\prime_3 
& = & \frac{i}{\sqrt{2+2 \frac{|\epsilon_{+}|}{|\epsilon_{-}|}}}
\left[e^{i(\phi_{-}-\phi_{+})/4}(\widetilde{S}^{\dag}+\widetilde{N}^{\dag}) 
- \sqrt{\frac{|\epsilon_{+}|}{|\epsilon_{-}|}} e^{-i(\phi_{-}-\phi_ {+})/4}
(\widetilde{S}+\widetilde{N}) \right] \nonumber \\ 
\widetilde{N}^\prime_4 
& = & \frac{1}{\sqrt{2+2 \frac{|\epsilon_{+}|}{|\epsilon_{-}|}}}
\left[e^{i(\phi_{-}-\phi_{+})/4}(\widetilde{S}^{\dag}+\widetilde{N}^{\dag}) 
+ \sqrt{\frac{|\epsilon_{+}|}{|\epsilon_{-}|}} e^{-i(\phi_{-}-\phi_ {+})/4}
(\widetilde{S}+\widetilde{N}) \right] \nonumber \\ 
\end{array}
\label{hamstates}
\end{equation}
and the eigenvalues 
\begin{equation}
\begin{array}{ccc}
\nu_1 & = & |M|-|\mu|/2-i \Gamma/2- e^{i(\phi_{-}+\phi_{+})/2}
\sqrt{|\epsilon_{+}|}\sqrt{|\epsilon_{-}|} \nonumber \\
\nu_2 & = & |M|-|\mu|/2-i \Gamma/2+ e^{i(\phi_{-}+\phi_{+})/2}
\sqrt{|\epsilon_{+}|}\sqrt{|\epsilon_{-}|} \nonumber \\
\nu_3 & = & |M|+|\mu|/2-i \Gamma/2- e^{i(\phi_{-}+\phi_{+})/2}
\sqrt{|\epsilon_{+}|}\sqrt{|\epsilon_{-}|} \nonumber \\
\nu_4 & = & |M|+|\mu|/2-i \Gamma/2+ e^{i(\phi_{-}+\phi_{+})/2}
\sqrt{|\epsilon_{+}|}\sqrt{|\epsilon_{-}|} \nonumber \\
\end{array}
\end{equation}
where
\begin{eqnarray}
\epsilon_{-} = |\epsilon_{-}| e^{i \phi_{-}} = |B_{SN}/(2 M)|- 
i \Gamma \epsilon_A/2 \nonumber \\
\epsilon_{+} = |\epsilon_{+}| e^{i \phi_{+}} = |B_{SN}/(2 M)|- 
i \Gamma \epsilon_A^*/2 \nonumber 
\end{eqnarray}

We consider an initial state at $t=0$ with equal number densities 
of the four sneutrino interaction states $\widetilde{F}_i$.
Using that the 
time evolution for the hamiltonian eigenstates Eq.(\ref{hamstates}) 
is trivially $\left|\widetilde{N}^\prime _i(t)\right>=
e^{-i\nu_it}\left|\widetilde{N}^\prime_i\right>$, 
we obtain that at time $t$ the interaction states are the
following:
\begin{eqnarray} 
\lefteqn{ \left|\widetilde{N}(t)\right>= {} } \nonumber  \\
& & {} = \frac{g_{1+}(t)+g_{2+}(t)}{4}~\left|\widetilde{N}\right> +
\sqrt{\frac{\epm}{\ep}}~ e^{i(\phim -\phip)/2}~ 
\frac{g_{1-}(t)-g_{2-}(t)}{4}~\left|\widetilde{N}^{\dag}\right> -   
{}   \nonumber  \\
& & {} -  \frac{g_{1+}(t)-g_{2+}(t)}{4}~\left|\widetilde{S}\right> -
\sqrt{\frac{\epm}{\ep}}~ e^{i(\phim -\phip)/2} ~  
\frac{g_{1-}(t)+g_{2-}(t)}{4}~\left|\widetilde{S}^{\dag}\right>
{} \nonumber  \\
\lefteqn{\left|\widetilde{N}^{\dag}(t)\right>= {} }  \nonumber \\
& & {} = \sqrt{\frac{\ep}{\epm}}~ e^{-i (\phim - \phip)/2}~ 
\frac{g_{1-}(t)-g_{2-}(t)}{4}~\left|\widetilde{N}\right>  +  
\frac{g_{1+}(t)+g_{2+}(t)}{4}~\left|\widetilde{N}^{\dag}\right>  
-   {}  \nonumber  \\
& & {} - \sqrt{\frac{\ep}{\epm}}~ e^{-i (\phim - \phip)/2} ~ 
\frac{g_{1-}(t)+g_{2-}(t)}{4}~\left|\widetilde{S}\right> - 
\frac{g_{1+}(t)-g_{2+}(t)}{4}~\left|\widetilde{S}^{\dag}\right>  
{}   \nonumber  \\
\lefteqn{\left|\widetilde{S}(t)\right>= {} } \nonumber  \\
& & {} = - \frac{g_{1+}(t)-g_{2+}(t)}{4}~\left|\widetilde{N}\right> -
\sqrt{\frac{\epm}{\ep}}~ e^{i(\phim -\phip)/2} ~
\frac{g_{1-}(t)+g_{2-}(t)}{4}~\left|\widetilde{N}^{\dag}\right> + {}
\nonumber \\
& & {} + \frac{g_{1+}(t)+g_{2+}(t)}{4}~\left|\widetilde{S}\right> +
\sqrt{\frac{\epm}{\ep}}~ e^{i(\phim -\phip)/2} ~
\frac{g_{1-}(t)-g_{2-}(t)}{4}~\left|\widetilde{S}^{\dag}\right> {}
\label{intereigenstates}\\
\lefteqn{\left|\widetilde{S}^{\dag}(t)\right>= {} } \nonumber  \\
& & {} = - \sqrt{\frac{\ep}{\epm}}~ e^{-i (\phim - \phip)/2}~
\frac{g_{1-}(t)+g_{2-}(t)}{4}~\left|\widetilde{N}\right> -
\frac{g_{1+}(t)-g_{2+}(t)}{4}~\left|\widetilde{N}^{\dag}\right> + {}
\nonumber \\
& & {}+ \sqrt{\frac{\ep}{\epm}}~ e^{-i (\phim - \phip)/2} ~
\frac{g_{1-}(t)-g_{2-}(t)}{4}~\left|\widetilde{S}\right> +
\frac{g_{1+}(t)+g_{2+}(t)}{4}~\left|\widetilde{S}^{\dag}\right>
\nonumber 
\end{eqnarray} 
The functions $g_{1\pm}(t)$ and $g_{2\pm}(t)$ containing the 
time dependence are given by:
\be \label{g1}
g_{1\pm}(t)=e^{-i\left(|M|-\frac{|\mu|}{2}-i\frac{\Gamma}{2}\right)t} 
\left[ e^{i \Delta \nu t}  \pm e^{-i \Delta \nu t} \right]~,
\ee
\be \label{g2}
g_{2\pm}(t)=e^{-i\left(|M|+\frac{|\mu|}{2}-i\frac{\Gamma}{2}\right)t} 
\left[ e^{i \Delta \nu t}  \pm e^{-i \Delta \nu t} \right]~,
\ee
with,
\be
\Delta \nu=e^{i\left(\frac{\phim+\phip}{2}\right)}~ \sqrt{\epm}~ \sqrt{\ep}~.
\ee

We neglect soft supersymmetry-breaking terms, so that 
$\widetilde{S}$ only decay 
to scalars, and $\widetilde{N}$ to antifermions:
\be \label{amplitudes}
\begin{array}{c}
|A[\widetilde{S}\to \widetilde{L}_k H]|^{2} =  
|A[\widetilde{S}^{\dag}\to \widetilde{L}_k^{\dag}H^{\dag}]|^{2}
=|Y_{1k}M|^2 \equiv |A_{\widetilde{L}_k}|^{2}  \\
|A[\widetilde{N}\to L^{\dag}_kh^{\dag}]|^{2} = 
|A[\widetilde{N}^{\dag}\to L_kh]|^{2}
=|Y_{1k}|^2 (s-m_L^2-m_h^2) \equiv |A_{L_k}|^{2}  \\
\end{array}
\ee

We next write the time dependent decay amplitudes in terms 
of $|A_{\widetilde{L}_k}|^2$ and $|A_{L_k}|^2$:
\bea  
\vspace{0.3cm}
|A[\widetilde{N}(t) \to  \widetilde{L}_k H]|^2 = 
|A[\widetilde{N}^{\dag}(t) \to  \widetilde{L}_k^{\dag}H^{\dag}]|^2 & =& 
\displaystyle{ \frac{|g_{1+}-g_{2+}|^2}{16}}~ |A_{\widetilde{L}_k}|^2  \\
\vspace{0.3cm}
|A[\widetilde{N}(t) \to  \widetilde{L}_k^{\dag}H^{\dag}]|^2  = 
\displaystyle{ \left( \frac{\epm}{\ep} \right)^2}
|A[\widetilde{N}^{\dag}(t) \to  \widetilde{L}_kH]|^2   & = &
\displaystyle{ \frac{\epm}{\ep}~ \frac{|g_{1-}+g_{2-}|^2}{16}}~ 
|A_{\widetilde{L}_k}|^2  \\
\vspace{0.3cm}
|A[\widetilde{N}(t) \to  L_kh]|^2  = 
\displaystyle{ \left( \frac{\epm}{\ep} \right)^2}
|A[\widetilde{N}^{\dag}(t) \to  L_k^{\dag}h^{\dag}]  & = & 
\displaystyle{\frac{\epm}{\ep}~ 
\frac{|g_{1-}-g_{2-}|^2}{16}}~ |A_{L_k}|^2  \\
\vspace{0.3cm}
|A[\widetilde{N}(t) \to  L_k^{\dag}h^{\dag}]|^2  = 
|A[\widetilde{N}^{\dag}(t) \to  L_kh]|^2  & = &
\displaystyle{\frac{|g_{1+}+g_{2+}|^2}{16}}~ |A_{L_k}|^2  \\
\vspace{0.3cm}
|A[\widetilde{S}(t) \to  \widetilde{L}_kH]|^2   = 
|A[\widetilde{S}^{\dag}(t) \to  \widetilde{L}_k^{\dag}H^{\dag}]|^2 & = & 
\displaystyle{\frac{|g_{1+}+g_{2+}|^2}{16} }~|A_{\widetilde{L}_k}|^2  \\
\vspace{0.3cm}
|A[\widetilde{S}(t) \to  \widetilde{L}_k^{\dag}H^{\dag}]|^2  = 
\displaystyle{ \left( \frac{\epm}{\ep} \right)^2}
|A[\widetilde{S}^{\dag}(t) \to  \widetilde{L}_kH]|^2 & = & 
\displaystyle{ \frac{\epm}{\ep}~ \frac{|g_{1-}-g_{2-}|^2}{16}}~ 
|A_{\widetilde{L}}|^2  \\
\vspace{0.3cm}
|A[\widetilde{S}(t) \to  L_kh]|^2  = 
\displaystyle{ \left( \frac{\epm}{\ep} \right)^2}
|A[\widetilde{S}^{\dag}(t) \to  L_k^{\dag}h^{\dag}]|^2 & = & 
\displaystyle{\frac{\epm}{\ep} 
~\frac{|g_{1-}+g_{2-}|^2}{16}}~ |A_{L_k}|^2  \\
\vspace{0.3cm}
|A[\widetilde{S}(t) \to  L_k^{\dag}h^{\dag}]|^2  =  
|A[\widetilde{S}^{\dag}(t) \to  L_kh]|^2  &= &
\displaystyle{\frac{|g_{1+}-g_{2+}|^2}{16}}~ |A_{L_k}|^2  
\eea
\label{amplitudes_temp}

We define the integrated $CP$ asymmetries for the fermionic and
scalar channels as:
\begin{eqnarray}
\epsilon_f&=&\frac 
{\int dt 
{\displaystyle \sum_{i,k}}\left[|A[\widetilde{F}_i(t)\to  L_k+X]|^2
-|A[\widetilde{F}_i(t)\to L_k^{\dag}+X]|^2\right]}
{\int dt 
{\displaystyle \sum_{i,k}}\left[|A[\widetilde{F}_i(t)\to  L_k+X]|^2
+|A[\widetilde{F}_i(t)\to L_k^{\dag}+X]|^2\right]}\nonumber \\
\epsilon_s&=&\frac 
{\int dt 
{\displaystyle \sum_{i,k}}\left[|A[\widetilde{F}_i(t)\to  \widetilde{L}_k+X]|^2
-|A[\widetilde{F}_i(t)\to \widetilde{L}_k^{\dag}+X]|^2\right]}
{\int dt 
{\displaystyle \sum_{i,k}}\left[|A[\widetilde{F}_i(t)\to  \widetilde{L}_k+X]|^2
+|A[\widetilde{F}_i(t)\to \widetilde{L}_k^{\dag}+X]|^2\right]}
\label{qmeps}
\end{eqnarray} 

Using the time-dependent amplitudes from Eq.~(\ref{amplitudes_temp}), 
the time integrated asymmetries are:
\be  
\label{epsilonl}
\epsilon_s=-\epsilon_f=\bar{\epsilon}=-\frac{1}{2}
\left( \frac{\epm}{\ep} - \frac{\ep}{\epm} \right) \chi~,
\ee
where the factor $\frac{\epm}{\ep} - \frac{\ep}{\epm}$ vanishes if $CP$ is 
conserved, and for $\epsilon_A \ll 1$ is given by:
\be 
\label{CPV} 
\frac{\epm}{\ep} - \frac{\ep}{\epm} \simeq 
\frac{4 M \Gamma |\epsilon_A|}{B_{SN}} \sin{\phi} = 
\frac{4 \Gamma\, A}{B_{SN}} \sin\phi
\ee
The time dependence is encoded in $\chi$:
\be  
\label{chi}
\chi=\frac{\int_{0}^{\infty} dt\left[ |g_{1-}(t)|^2+|g_{2-}(t)|^2\right]}
{\int_{0}^{\infty} dt\left[ |g_{1+}(t)|^2 + |g_{2+}(t)|^{2} +
|g_{1-}(t)|^2+|g_{2-}(t)|^2 \right]}~.
\ee

In the limit $\epsilon_A \ll 1$ the time integrals are :
\be
\int_{0}^{\infty} dt\left[ |g_{1-}|^2+|g_{2-}|^2\right] 
\simeq 4 ~ \frac{|B_{SN}|^2/\Gamma |M|^2}{\Gamma^2+|B_{SN}|^2/|M|^2}~,
\ee
\be
\int_{0}^{\infty} dt\left[ |g_{1+}|^2+|g_{2+}|^2 +
|g_{1-}(t)|^2+|g_{2-}(t)|^2 \right] 
\simeq  \frac{8}{\Gamma}~;
\ee
\be
\epsilon_s=-\epsilon_f = \bar{\epsilon}=-
\frac{\Gamma ~ |B_{SN}| \,| A|}{\Gamma^2 |M|^2+|B_{SN}|^2} ~ \sin\phi
\label{qmasym}
\ee

The comparison between the asymmetry in Eq.(\ref{qftasym}) and 
Eq.(\ref{qmasym}) is in full analogy to the corresponding comparison 
in the standard see-saw case discussed in \cite{soft1}. 
The asymmetry computed in the quantum mechanics approach, 
based on an effective (non hermitic) Hamiltonian, Eq.(\ref{qmasym}),
agrees with the one obtained using a field-theoretical approach, 
Eq.(\ref{qftasym}) in the limit $\Gamma \ll B_{SN}/M$.
When $\Gamma \gg B_{SN}/M$, the four sneutrino states become
two pairs of not well-separated  particles. In this case 
the result for the asymmetry can depend on how the
initial state is prepared. If one assumes  that 
the singlet sneutrinos are in a thermal 
bath with a thermalization time $\Gamma^{-1}$ shorter than the 
typical oscillation times, $\Delta M_{ij}^{-1}$, coherence 
is lost and it is appropriate to compute the CP asymmetry  in terms of the 
mass eigenstates Eq.(\ref{masseigenstates}) 
as done in Sec.~\ref{qft}
and one obtains Eq.~(\ref{qftasym}). 
If, on the contrary, one assumed that the $\tilde N, \tilde S$ states 
are produced in interaction eigenstates and coherence is not lost in their
evolution, then it is appropriate to compute the CP asymmetry in
terms of the interaction eigenstates Eq.(\ref{intereigenstates})
as done in this appendix and one obtains Eq.(\ref{qmasym}).


\begin{thebibliography}{10}
%
\bibitem{fy}
M. Fukugita and T. Yanagida, Phys. Lett. {\bf B174} (1986) 45
%
\bibitem{ss} 
P.~Minkowski,
  Phys.\ Lett.\ B {\bf 67}, 421 (1977);
M. Gell-Mann, P. Ramond and R. Slansky, Proceedings of   
the Supergravity Stony Brook Workshop, New York, 1979, eds. P. Van   
Nieuwenhuizen and D. Freedman (North-Holland, Amsterdam);
T. Yanagida, Proceedings of
the  Workshop  on Unified  Theories  and  Baryon  Number in the  
Universe,  Tsukuba,  Japan 1979 (eds. A.  Sawada and A.
Sugamoto, KEK Report No.  79-18, Tsukuba); 
R.~Mohapatra and G.~Senjanovic, 
Phys.\ Rev.\ Lett.\ {\bf 44}, 912 (1980).
%
\bibitem{di}
  S.~Davidson and A.~Ibarra,
  Phys.\ Lett.\ B {\bf 535} (2002) 25
  [arXiv:hep-ph/0202239].
%
\bibitem{Mbound}
W.~Buchmuller, P.~Di Bari and M.~Plumacher,
Nucl.\ Phys.\ B {\bf 643} (2002) 367 [arXiv:hep-ph/0205349];
J.~R.~Ellis and M.~Raidal,
Nucl.\ Phys.\ B {\bf 643} (2002) 229 [arXiv:hep-ph/0206174].
%
\bibitem{flavour}
  A.~Abada, S.~Davidson, F.~X.~Josse-Michaux, M.~Losada and A.~Riotto,
  JCAP {\bf 0604} (2006) 004
  [arXiv:hep-ph/0601083];
 E.~Nardi, Y.~Nir, E.~Roulet and J.~Racker,
  JHEP {\bf 0601}, 164 (2006)
  [arXiv:hep-ph/0601084];
 A.~Abada, S.~Davidson, A.~Ibarra, F.~X.~Josse-Michaux, M.~Losada and 
 A.~Riotto,
  arXiv:hep-ph/0605281;
 S.~Blanchet and P.~Di Bari,
  arXiv:hep-ph/0607330.
 
  %
%
%
\bibitem{ma}
 E.~Ma, N.~Sahu and U.~Sarkar,
  J.\ Phys.\ G {\bf 32}, L65 (2006)

\bibitem{db1}
 P.~Di Bari,
  Nucl.\ Phys.\ B {\bf 727} (2005) 318
  [arXiv:hep-ph/0502082].

\bibitem{oscar}
 O.~Vives,
  Phys.\ Rev.\ D {\bf 73} (2006) 073006
  [arXiv:hep-ph/0512160].

%
\bibitem{gravi} 
M.~Y.~Khlopov and A.~D.~Linde,
Phys.\ Lett.\ B \textbf{138} (1984) 265;
J.~R.~Ellis, J.~E.~Kim and D.~V.~Nanopoulos,
Phys.\ Lett.\ B \textbf{145} (1984) 181;
J.~R.~Ellis, D.~V.~Nanopoulos and S.~Sarkar,
Nucl.\ Phys.\ B \textbf{259} (1985) 175;
T.~Moroi, H.~Murayama and M.~Yamaguchi,
Phys.\ Lett.\ B \textbf{303} (1993) 289;
M.~Kawasaki, K.~Kohri and T.~Moroi,
Phys.\ Lett.\ B {\bf 625} (2005) 7;
For a recent discussion, see:
  K.~Kohri, T.~Moroi and A.~Yotsuyanagi,
  Phys.\ Rev.\ D {\bf 73} (2006) 123511
%
\bibitem{soft1}
Y.~Grossman, T.~Kashti, Y.~Nir and E.~Roulet,
  Phys.\ Rev.\ Lett.\  {\bf 91} (2003) 251801
  [arXiv:hep-ph/0307081];
  JHEP {\bf 0411} (2004) 080
  [arXiv:hep-ph/0407063].
%
\bibitem{soft2}
  G.~D'Ambrosio, G.~F.~Giudice and M.~Raidal,
  Phys.\ Lett.\ B {\bf 575}, 75 (2003)
  [arXiv:hep-ph/0308031].
%
\bibitem{anna}
G.~D'Ambrosio, T.~Hambye, A.~Hektor, M.~Raidal and A.~Rossi,
  Phys.\ Lett.\ B {\bf 604} (2004) 199
  [arXiv:hep-ph/0407312].
%
\bibitem{iss}
R.N. Mohapatra and J.W.F. Valle, Phys. Rev. {\bf D34} (1986) 1642 
%
\bibitem{tHooft}
G. 't Hooft, Lecture given at Cargese Summer Inst., Cargese, France, 
Aug. 26 - Sep. 8, 1979.
%
\bibitem{pheno1}
J.~Bernabeu, A.~Santamaria, J.~Vidal, A.~Mendez and J.~W.~F.~Valle,
  Phys.\ Lett.\ B {\bf 187} (1987) 303;
 G.~C.~Branco, M.~N.~Rebelo and J.~W.~F.~Valle,
  Phys.\ Lett.\ B {\bf 225} (1989) 385;
 N.~Rius and J.~W.~F.~Valle,
  Phys.\ Lett.\ B {\bf 246} (1990) 249.
 M.~C.~Gonzalez-Garcia and J.~W.~F.~Valle,
  Mod.\ Phys.\ Lett.\ A {\bf 7} (1992) 477.
 D.~Tommasini, G.~Barenboim, J.~Bernabeu and C.~Jarlskog,
  Nucl.\ Phys.\ B {\bf 444} (1995) 451
  [arXiv:hep-ph/9503228].
 F.~Deppisch, T.~S.~Kosmas and J.~W.~F.~Valle,
  Nucl.\ Phys.\ B {\bf 752} (2006) 80
  [arXiv:hep-ph/0512360].
 %
\bibitem{pheno2}
 M.~Dittmar, A.~Santamaria, M.~C.~Gonzalez-Garcia and J.~W.~F.~Valle,
  Nucl.\ Phys.\ B {\bf 332} (1990) 1.
  M.~C.~Gonzalez-Garcia, A.~Santamaria and J.~W.~F.~Valle,
  Nucl.\ Phys.\ B {\bf 342} (1990) 108.
%
\bibitem{hr}
 P.~Hernandez and N.~Rius,
  Nucl.\ Phys.\ B {\bf 495}, 57 (1997)
  [arXiv:hep-ph/9611227].
%
 \bibitem{pheno3}
 E.~Nardi, E.~Roulet and D.~Tommasini,
  Phys.\ Lett.\ B {\bf 327} (1994) 319
  [arXiv:hep-ph/9402224].
 E.~Nardi, E.~Roulet and D.~Tommasini,
  Phys.\ Lett.\ B {\bf 344} (1995) 225
  [arXiv:hep-ph/9409310].
%
\bibitem{alsv}
E.~Akhmedov, M.~Lindner, E.~Schnapka and J.~W.~F.~Valle,
  Phys.\ Rev.\ D {\bf 53} (1996) 2752
  [arXiv:hep-ph/9509255];
  Phys.\ Lett.\ B {\bf 368} (1996) 270
  [arXiv:hep-ph/9507275].
%
\bibitem{barr}
 S.~M.~Barr,
  Phys.\ Rev.\ Lett.\  {\bf 92} (2004) 101601
  [arXiv:hep-ph/0309152];
 S.~M.~Barr and I.~Dorsner,
  Phys.\ Lett.\ B {\bf 632} (2006) 527
  [arXiv:hep-ph/0507067].
%
\bibitem{cv1}
 M.~C.~Gonzalez-Garcia and J.~W.~F.~Valle,
  Phys.\ Lett.\ B {\bf 216} (1989) 360.
%
\bibitem{siss}
 F.~Deppisch and J.~W.~F.~Valle,
  Phys.\ Rev.\ D {\bf 72} (2005) 036001
  [arXiv:hep-ph/0406040].
%
\bibitem{farzan}
 Y.~Farzan,
  Phys.\ Rev.\ D {\bf 69}, 073009 (2004).
%
\bibitem{kolb}
E.~W.~Kolb and S.~Wolfram,
  Nucl.\ Phys.\ B {\bf 172}, 224 (1980)
  [Erratum-ibid.\ B {\bf 195}, 542 (1982)].
%
\bibitem{pi}
A. Pilaftsis, Phys. Rev. {\bf D56} (1997) 5431 [arXiv:hep-ph/9707235] 
%
\bibitem{thermal}
 G.~F.~Giudice, A.~Notari, M.~Raidal, A.~Riotto and A.~Strumia,
  Nucl.\ Phys.\ B {\bf 685} (2004) 89
  [arXiv:hep-ph/0310123].
%
\bibitem{WMAP}
 D.~N.~Spergel {\it et al.},
  arXiv:astro-ph/0603449.
\bibitem{king}
  S.~F.~King,
  Nucl.\ Phys.\ B {\bf 576} (2000) 85
%
\bibitem{grossman}
  Y.~Grossman, R.~Kitano and H.~Murayama,
  JHEP {\bf 0506}, 058 (2005).
%
\bibitem{wagner}
A.~D.~Medina and C.~E.~M.~Wagner,
  arXiv:hep-ph/0609052.
\end{thebibliography}
\end{document}